\documentclass[12pt,english]{article}
\usepackage[T1]{fontenc}
\usepackage[latin1]{inputenc}
\usepackage{a4wide}
\usepackage{babel}

\makeatletter

\providecommand{\LyX}{L\kern-.1667em\lower.25em\hbox{Y}\kern-.125emX\@}

\usepackage[T1]{fontenc}
\usepackage{a4wide}
\usepackage{babel}
\usepackage{color}
\usepackage{setspace}
\makeatletter

\usepackage[T1]{fontenc}

\makeatletter

\usepackage[T1]{fontenc}
\usepackage{a4}



\makeatother

\makeatother
\begin{document}
{\raggedleft December 2001\par}
{\raggedleft Revised version\bigskip{}\par}
{\raggedleft KIAS-P01033\par}
{\raggedleft BROWN-HET-1278\par}
{\raggedleft LPNHE 2001-08\bigskip{}\par}

{\centering {\large Hadronic scattering amplitudes: }\large \par}

{\centering {\large medium-energy constraints on asymptotic behaviour}
\bigskip{}\par}

{\centering J.R. Cudell\( ^{a} \), V.V. Ezhela\( ^{b} \), P. Gauron\( ^{c} \),
K. Kang\( ^{d} \), Yu.V. Kuyanov\( ^{b} \), S.B. Lugovsky\( ^{b} \),
\\
B. Nicolescu\( ^{c} \), and N.P. Tkachenko\( ^{b} \)\par}

\medskip{}
{\centering COMPETE\footnote{%
COmputerised Models, Parameter Evaluation for Theory and Experiment.
} collaboration\par}
\bigskip{}

\noindent {\footnotesize \( ^{a} \)Institut de Physique, Bât. B5,
Université de Liège, Sart Tilman, B4000 Liège, Belgium}{\footnotesize \par}

\noindent {\footnotesize \( ^{b} \)COMPAS group, IHEP, Protvino,
Russia}{\footnotesize \par}

\noindent {\footnotesize \( ^{c} \)LPNHE}\footnote{%
Unité de Recherche des Universités Paris 6 et Paris 7, Associé au
CNRS
}{\footnotesize -LPTPE, Université Pierre et Marie Curie, Tour 12 E3,
4 Place Jussieu, 75252 Paris Cedex 05, France}{\footnotesize \par}

\noindent {\footnotesize \( ^{d} \)Physics Department, Brown University,
Providence, RI, U.S.A.}{\footnotesize \par}
\bigskip{}

\begin{abstract}
We consider several classes of analytic parametrisations of hadronic
scattering amplitudes, and compare their predictions to all available
forward data (\( pp \), \( \overline{p}p \), \( \pi p \), \( Kp \),
\( \gamma p \), \( \gamma \gamma  \), \( \Sigma p \)). Although
these parametrisations are very close for \( \sqrt{s}\geq 9 \) GeV,
it turns out that they differ markedly at low energy, where a universal
pomeron term \( \sim \ln ^{2}s \) enables one to extend the fit
down to \( \sqrt{s} \)=4 GeV.
\end{abstract}

\section{Introduction}

The singularity structure of forward hadronic amplitudes is of great
importance, as it controls the extrapolation of cross sections to
high energies and to small \( x \). Its study lies mostly outside
the realm of perturbative QCD, except perhaps at small \( x \) and
high \( Q^{2} \), where there is some overlap, hence the hope to
obtain some QCD-based understanding of these amplitudes in the near
future. However, there are several tools available to treat this non
perturbative domain. These are based on the theory of the analytic
S matrix.

The first is to demand that hadronic amplitudes are analytic functions
in the complex angular momentum \( J \). The singularities in the
complex \( J \) plane then determine the form of the asymptotic amplitudes
in \( s \) at finite \( t \). This means that one can then relate,
through analyticity and crossing symmetry, the real part of the amplitude
to its imaginary part. In other words, the exact knowledge of the
cross section for all \( s \) is equivalent to that of the \( \rho  \)
parameter. In practice, there are several analytic forms which are
very close for the total cross sections in a finite interval in \( s \),
but which differ markedly for the real part. Hence in this paper,
we shall consider the experimental constraints on both the real and
the imaginary parts. Furthermore, \( t \)-channel unitarity leads
to the conclusion that these singularities should be universal, in
the sense that they do not depend on the scattering hadrons\footnote{%
The photon is special in this context, and may have further singularities.
}. This leads to factorizing amplitudes\footnote{%
Note however that factorization can be proven only for simple poles.
}, for which the residue depends on the colliding hadrons, but the
singularity is independent of them.

The second constraint is due to the unitarity of partial waves and
polynomial boundedness of the absorptive part within the Lehmann ellipse.
This leads to the celebrated Froissart-Martin bound \cite{Froissart},
which indicates that at asymptotic energies, total cross sections
cannot increase faster than \( \ln ^{2}s \) (note that this behaviour
was first proposed by W. Heisenberg in 1952 \cite{Heisenberg:1952}).
Although this is a priori a strong constraint, it turns out that the
coefficient of the \( \ln ^{2}s \) can be large: all we know is
that it is bounded by\break \( \frac{\pi }{m_{\pi }^{2}}\simeq 60 \)
mb (Lukaszuk-Martin \cite{Martin}), hence parametrisations which
asymptotically violate the Froissart bound, such as rising simple
poles, may survive to present energies without violating unitarity.

Finally, the last ingredient is Regge theory. The meson trajectories
can indeed be seen in a Chew-Frautschi plot, and hence their intercepts
can in principle be measured directly. This leads to the conclusion
that the intercepts of these trajectories are of order 0.5, that the
\( C=+1 \) and \( C=-1 \) trajectories are approximately degenerate,
and that they seem to be linear. We shall assume in the following
that their contribution to the total cross section can be parametrized
by \( Y^{+}s^{\alpha _{+}-1} \) and \( Y^{-}s^{\alpha _{-}-1} \).

These constraints, unfortunately, are far from providing a unique
answer. As an example, the derivative relations \cite{Kang:1975gt}
can be conceived as a source of an infinite class of analytic parametrisations
satisfying the above theoretical criteria. However, it is possible
to reduce this class of models to a few exemplar cases, for which
the cross section, in the limit \( s\rightarrow \infty  \), behaves
as a constant, as \( \ln s \) or as \( \ln ^{2}s \). Hence in
practice, only a handful of parametrisations have been considered
and constrained. These represent variations on the parametrisation
proposed in \cite{Amaldi:1980kd, log parametrisations}, which will
be symbolically referred to as (Regge + Regge + Pomeranchuk + Heisenberg)
type parametrisations -- \( {\textrm{RRPH}} \).
Here both \( {\textrm{R}} \) stand for the leading reggeon terms,
\( {\textrm{P}} \) stands for a constant contribution to the total
cross section at asymptotic energies (the classical Pomeranchuk asymptotic
limit \cite{Pom}) and \( {\textrm{H}} \) stands for the asymptotically
infinitely rising with energy contribution, which we take as \( \ln s \)
or \( \ln ^{2}s \). Because of its popularity and simplicity,
we shall also consider case E, i.e. the case of a simple pole \( s^{\alpha
_{\wp }-1} \)with
\( \alpha _{\wp }>1 \).

Some of us (COMPAS) are maintaining a complete set of data for all
hadronic processes, so that we are in a position to fully evaluate
the various possibilities. We are using a slightly improved dataset
from the one of \cite{Cudell:2000tx}: some preliminary data on the
\( \rho  \) parameter have been removed, and new published data from
SELEX (\( {\pi }^{-}N \) and \( \Sigma ^{-}N \) at 600 GeV/c)
\cite{Dersch:2000zg}
and OPAL (\( \gamma \gamma  \)) \cite{Abbiendi:2000sz} were added.
We did not use the new recent data from L3 \cite{Acciarri} on $\gamma \gamma
\to hadrons$
total hadronic cross sections because unfortunately these very interesting
data are still not published yet. Definitely these data, when published,
will be used in the next iteration of the cross assessments.

In the past few years, and mainly because of the existence of this
dataset, several advances have been made:

\begin{enumerate}
\item The systematic and simultaneous study, via analytic representations,
of the forward data, both \( \sigma _{tot} \) and \( \rho  \), for
\( pp \), \( \overline{p}p \), \( \pi ^{\pm }p \), \( K^{\pm }p \),
\( \gamma p \) and \( \gamma \gamma  \) scattering. Such a program
was initiated by the COMPAS group\cite{Groom:2000in}, and pursued
in refs. \cite{Cudell:2000tx,Landshoff:2000mu};
\item The general recognition that a Regge pole model \cite{Landshoff:2000mu}
has a much wider range of applicability than previously expected while
it was also recognized that the exchange-degenerate reggeons were
not preferred by the forward scattering data \cite{CKK};
\item The rediscovery \cite{Gauron:2000ri} of former ideas \cite{solovev:1973,
Ter-Martirosian:1988yy}
such as a 2-component soft pomeron, with one component taking quark
counting into account and the other being universal and rising with
energy, or of full lifting of degeneracy for lower meson trajectories
\cite{Desgrolard}.
\item The impossibility to distinguish between wide ranges of analytic
parametrisations
when using data at \( \sqrt{s}\geq 9 \) GeV \cite{Cudell:2000tx}.
\end{enumerate}
We want to examine in detail those conclusions, and see to which extents
the models considered in \cite{Cudell:2000tx} can be extended to
lower energy, i.e. above the resonance region\break \( \sim 3 \)
GeV. {A new quantitative procedure of ranking models by the quality
of the fit to the current experimental data is suggested and used.
In section 2, we shall concentrate on total cross sections, and propose
this new ranking scheme. In section 3, we shall extend our analysis
to all forward data, and see that this changes the picture considerably.
In section 4, we shall comment on some models proposed recently, and
which were not considered directly in the previous analysis. In section
5, we shall comment on cosmic ray data. To conclude, we shall present
the possible alternatives, and analyse their respective drawbacks
and advantages.}

\section{Fits to lower-energy total cross sections}

As it will turn out, the consideration of \( \rho (s) \) data results
in a very constrained fit, but some of the sub-samples of data are
poorly fitted to. This might be blamed on the quality and systematic
errors on the forward-scattering data for \( \rho (s) \). Hence the
first and safest constraint must be the reproduction of \( \sigma _{tot}(s) \)
data only. In this case, the number of possible models that achieve
a good \( \chi ^{2} \) per degree of freedom (\( \chi ^{2}/dof \))
is quite large. To describe the different possibilities we will need
some notations to classify variants, and we shall use the following:

\begin{equation}
\label{SIG}
\sigma _{}^{a_{\mp }b}=\frac{1}{s}\left( {\textrm{R}}^{+ab}(s)\pm
{\textrm{R}}^{-ab}(s)+{\textrm{P}}^{ab}+{\textrm{H}}^{ab}(s)\right) ,
\end{equation}

\noindent \vspace*{3mm} where: 
\begin{itemize}
\item ${\textrm{R}}^{+ab}(s)=Y^{ab}_{1}\cdot (s/s_{1})^{\alpha _{1}}$,
with \( s_{1}=1 \) GeV\( ^{2} \), 
\item ${\textrm{R}}^{-ab}(s)=Y^{ab}_{2}\cdot (s/s_{1})^{\alpha _{2}}$,
\item ${\textrm{P}}^{ab}=sC^{ab}$ is the Pomeron simple pole at $J = 1$
\item \( {\textrm{H}}^{ab}(s) \)
stands for one of the three following possibilities : 
\subitem . a supplementary simple pole at $J=\alpha _{\wp}$, with $\alpha _{\wp}>1$:
$$ {\textrm{E}}^{ab}=X^{ab}(s/s_{1})^{\alpha _{\wp }}; $$
\subitem . a double pole at $J=1$: 
$$L_{ab} = s (B_{ab} \ln(s/s_1) + A_{ab});$$
\subitem . a triple pole at $J=1$:
$$L2_{ab} = s (B_{ab} (\ln^2(s/s_0) + A_{ab}).$$
\end{itemize}
In the general case, the constants $C_{ab}$ and $A_{ab}$ are independent 
and they
are associated with a different behaviour in $t$. But at $t = 0$, as is the
case for our fits, they can't be distinguished. They mix and we are
left, when we consider logarithms, with just linear or quadratic forms 
in $\ln s$:
$$P_{ab} + L_{ab} =s B_{ab} \ln(s/s_1) + sZ_{ab}$$
and 
$$P_{ab} + L2_{ab} = s B_{ab} \ln^2(s/s_0)  + s Z_{ab},$$
where 
$Z_{ab} = C_{ab} + A_{ab}$.

In the following we will restrict ourselves to fits where $s_0$ is
process-independent. We have also considered fits with the ratio $Z_{ab}/B_{ab}$\ \ kept
process-independent
$$P_{ab} + L_{ab} = s \lambda_{ab} (B \ln(s/s_1) + A),$$
with $\lambda_{pp} = 1$, as well as fits to the form RRE, 
without any P term.

We fit to 3 pairs of reactions for particle and antiparticle: $pp$ and $\bar p p$, $\pi^\pm p$ and $K^\pm p$, one reaction with particles $\Sigma^- p$
and two reactions coupled only to C=+1 trajectories: $\gamma p$ and $\gamma\gamma$. \\

The counting of parameters then goes as follows:
\begin{itemize}
\item one intercept, and 6 residues ({\it i.e.} 7 parameters) for each $C=+1$ reggeon;
\item one intercept and 4 residues ({\it i.e.} 5 parameters) for each $C=-1$ reggeon.
\end{itemize}
Concerning the pomeron terms, unless otherwise indicated by the subscript $nf$, we impose factorisation 
of the $\gamma$ cross sections:
$H_{\gamma \gamma}=\delta H_{\gamma p}=\delta^2 H_{pp}$ and/or
$P_{\gamma \gamma}
=\delta P_{\gamma p}=\delta^2 P_{pp}$ with the same value of $\delta$. 
This leads to: 
\begin{itemize}
\item 1 parameter $\delta$;
\item 4 parameters for the constant term $Z_{ab}$;
\item 4 parameters $B_{ab}$ + one intercept for E or one scale factor
$s_0$ for L2;
\end{itemize}
When considering several singularities for the pomeron term, we usually treat 
them as independent. However, when we implement factorisation, we take the
same value of $\delta$ for all singularities. 
This leads to:
\begin{itemize}
\item 9 parameters for PL;
\item 10 parameters for PL2 or PE.
\end{itemize}

Furthermore, we have considered several possibilities to constrain
the parameters. The following notations are attached as either superscript
or subscript to the model variants in each case:
\begin{description}
\item [\emph{d}]means degenerate leading reggeon trajectories \( \alpha
_{1}=\alpha _{2} \). This lowers the number of parameters by 2 units, as
one has only one intercept, and one coupling for the $\Sigma^-p$ cross section;
\item [\emph{u}]means universal for the rising term 
(independent of projectile hadron). This reduces
the number of parameters by 3 units. Assuming again the same factorisation
for all pomeron singularities, we get 6 parameters for PL$_u$, and 7 for PL2$_u$;
\item [\emph{nf}]means that we have not imposed factorization for the residues
of \( {\textrm{H}}^{ab}(s) \) in the case of the \( \gamma \gamma  \)
and \( \gamma p \) cross sections. This adds one parameter to the fit
in the case of a single pomeron singularity, and two or three for multiple singularities.
\item [\emph{qc}] means that approximate quark counting rules of the
additive quark model \cite{Levin:1965mi} are imposed on the residues. This means
that the u, d and s couplings can be deduced from $pp$, $\pi p$ and $K p$
scattering, and used to predict $\Sigma p$. Hence this lowers the number
of parameters by 1 unit per singularity to which this rule is applied.
It should 
be noted that analogous counting rules also follow from the so-called 
gluon dominance model \cite{Kiselev:1986fr} for the dominant asymptotic contribution to the cross
sections. These counting rules were confirmed to some extent recently 
in the global fits of \cite{Cudell:2000tx}. 
\end{description}
Finally, we have sometimes assumed that the ratio of the residues of
different singularities is process-independent.  
This is noted by including these singularities in braces \{\}.
We have also considered the possibility that factorisation works for 
the lower $C=+1$ trajectories, with the same $\delta$ as for the pomeron.
We indicate this by putting the singularities in brackets [].

All reasonable combinations of these constraints give more than 256
different variants of the parametrisations. We shall consider here
only seven representative models that give a \( \chi ^{2}/dof \)
smaller than 1.5 for all considered energies. Further results may
be found in Appendices 1 and 2.

Table 1 gives the results for the minimum center-of-mass energy considered
in the fit 
$ \sqrt{s_{min}}= 3 $ GeV. Note that
because of the large number of points, slight deviations of the \( \chi
^{2}/dof \)
from 1 result in a very low confidence level. Hence we have shown
the area of applicability of the models as the energy values for which
\( \chi ^{2}/dof\leq 1.0 \).
\bigskip{}

{\centering {\small \begin{tabular}{|l|c|c|c|c|c|c|c|c|}
\cline{2-9}
\multicolumn{1}{l|}{}&
\multicolumn{8}{c|}{ {\small \( \sqrt{s_{min}} \) in GeV (number of points)
}}\\
\hline
\multicolumn{1}{|l|}{ {\small \( {\textrm{Model code}\, (\textrm{N}_{par})} \)
}}&
{\small 3 (725)}&
{\small 4 (580)}&
{\small 5 (506) }&
{\small 6 (433) }&
{\small 7 (368) }&
{\small 8 (330) }&
{\small 9 (284) }&
{\small 10 (229) }\\
\hline
{\small \( {\textrm{RRE}_{nf}(19)} \)}&
{\small 1.38 }&
{\small 1.15 }&
{\small \( \bf 0.91 \)}&
{\small \( \bf 0.87 \)}&
{\small \( \bf 0.89 \)}&
{\small \( \bf 0.90 \)}&
{\small \( \bf 0.93 \)}&
{\small \( \bf 0.91 \)}\\
\hline
{\small \( {\textrm{RRE}^{qc}}(17) \)}&
{\small 1.39 }&
{\small 1.17 }&
{\small \( \bf 0.93 \)}&
{\small \( \bf 0.89 \)}&
{\small \( \bf 0.90 \)}&
{\small \( \bf 0.91 \)}&
{\small \( \bf 0.93 \)}&
{\small \( \bf 0.92 \)}\\
\hline
{\small \( {\textrm{RRL}_{nf}(19)} \)}&
{\small 1.31 }&
{\small \( \bf 0.96 \)}&
{\small \( \bf 0.82 \)}&
{\small \( \bf 0.80 \)}&
{\small \( \bf 0.85 \)}&
{\small \( \bf 0.85 \)}&
{\small \( \bf 0.86 \)}&
{\small \( \bf 0.85 \)}\\
\hline
{\small \( {\textrm{RRPL}}(21) \)}&
{\small 1.33 }&
{\small \( \bf 0.98 \)}&
{\small \( \bf 0.85 \)}&
{\small \( \bf 0.83 \)}&
{\small \( \bf 0.87 \)}&
{\small \( \bf 0.88 \)}&
{\small \( \bf 0.84 \)}&
{\small \( \bf 0.74 \)}\\
\hline
{\small \( {(\textrm{RR})^{d}\textrm{ P$_{nf}$ L}2}(20) \)}&
{\small 1.24 }&
{\small \( \bf 0.99 \)}&
{\small \( \bf 0.82 \)}&
{\small \( \bf 0.79 \)}&
{\small \( \bf 0.83 \)}&
{\small \( \bf 0.84 \)}&
{\small \( \bf 0.83 \)}&
{\small \( \bf 0.73 \)}\\
\hline
{\small \( {\textrm{RRP$_{nf}$ L}2_{u}}(21) \)}&
{\small 1.26 }&
{\small \( \bf 0.97 \)}&
{\small \( \bf 0.81 \)}&
{\small \( \bf 0.79 \)}&
{\small \( \bf 0.82 \)}&
{\small \( \bf 0.83 \)}&
{\small \( \bf 0.82 \)}&
{\small \( \bf 0.75 \)}\\
\hline
{\small \( {(\textrm{RR})^{d}\textrm{ P L}2_{u}}(17) \)}&
{\small 1.28 }&
\textbf{\small 1.0} {\small }&
{\small \( \bf 0.82 \)}&
{\small \( \bf 0.81 \)}&
{\small \( \bf 0.83 \)}&
{\small \( \bf 0.83 \)}&
{\small \( \bf 0.83 \)}&
{\small \( \bf 0.76 \) }\\
\hline
\end{tabular}}\small \par}

\begin{quote}
\medskip{} {\small Table 1: the \( \chi ^{2}/dof \) of best models
fitting all cross section data down to 4 GeV. Numbers in bold represent
the area of applicability of each model. In parenthesis, we indicate
the number of parameters ($N_{par}$) for each model.}{\small \par}
\end{quote}
\bigskip{}
As can be seen, the data are compatible with many possibilities, and
one cannot decide at this level what the nature of the pomeron is,
and whether any of the regularities considered above is realised.
Note that 9 (resp. 23) models shown in Appendix 1 fit the data well ({\it
i.e.} with a $\chi^2/dof<1$)
for $ \sqrt{s_{min}}= 4 $ GeV (resp. 5 GeV). Hence it seems that sub-leading
trajectories and other non-asymptotic characteristics do not manifest
themselves yet. One can see that the logarithmic increases in general
fit better than simple powers, even at large energy \( \sqrt{s}\sim 10 \)
GeV, but that the difference in \( \chi ^{2}/dof \) is not large
enough to reach any firm conclusion. Quark counting can be implemented
for each possible rising term, but on the other hand one can choose
a universal (beam-independent) rise as well. It is interesting that
a reasonable degeneracy of the leading reggeon trajectories can be
implemented only in models which have a \( \ln ^{2}s \) pomeron.
The latter degeneracy is in fact expected to hold in global fits to
the forward scattering data of all hadronic processes, when one includes
$K^{+}p$ scattering, which has an exotic s-channel in view of
duality.

We can choose two approaches to distinguish further amongst the above
models. We can add more data, which we shall do in the next section,
but we want first to examine in detail the quality of the fits. Indeed,
despite the fact that these models do fit the data well \emph{globally,}
several other characteristics may be considered, and demanded on the
results. We shall present here a set of indicators which quantify
several aspects of the fits, and which will enable us to assess better
the quality of the models.

\subsection{Indicators measuring the quality of the fits.}

The best known such quantity is certainly the \( \chi ^{2}/dof \),
or more precisely the \emph{}confidence level (CL).

However, because Regge theory does not apply in the resonance region,
no model is expected to reproduce the data down to the lowest measured
energy. The cutoff we have given in Table 1 is \emph{ad hoc}: we know
the fits must fail at some point, but we cannot predict where. Hence
another indicator will be the \emph{range} of energy of the data that
the model can reproduce with a \( \chi ^{2}/dof\leq 1.0 \).

Furthermore, the quality of the data varies depending on which quantity
or which process one considers. In principle, one could introduce
some kind of data selection, but that would undoubtedly bias the fits
one way or the other. The other option is to assign a weight to each
process or quantity, which takes into account the quality of the data.
Given that this will be done to compare models together, we are certainly
entitled to choose the weights as determined by the best fit. Hence
we introduce \[
w_{j}=\min \left( 1,\frac{1}{\chi _{j}^{2}/nop}\right) \]
 where \( j=1, \ldots \) 9 refers to the process, and we define the renormalised
\( \chi _{R}^{2} \) as:

\[
\chi ^{2}_{R}\equiv \sum _{j}w_{j}\chi ^{2}_{j}\]

Finally, if a fit is physical in a given range, then its parameters
must be stable if one considers part of the range: different determinations
based on a sub-sample must be compatible. Hence another indicator
will deal with the \emph{stability} of the fit.

We have developed a series of statistical quantities that enable us
to measure the above features of the fits. All these indicators are
constructed so that the higher their value the better is the quality
of the data description.

\begin{description}
\item [(1)]\textbf{The} \textbf{Applicability Indicator:} It characterizes
the range of energy which can be fitted by the model. This range can
in principle be process-dependent, but we shall not consider such
a case here. The range of applicability is, by definition, the range
of data where fit has a confidence level bigger CL \( >50\% \). One
of the simplest variant is as follows: \begin{equation}
\label{A}
A_{j}^{M}=w_{j}\ln (E^{M,high}_{j}/E^{M,low}_{j}),\qquad A^{M}={1\over
N_{sets}}\sum _{j}A^{M}_{j}
\end{equation}
 where \( j \) is the multi-index {denoting} the pair (data subset,
observable); $N_{sets}$ is the number of such subsets, 
\( E^{M,high}_{j} \) is the highest value of the energy
in the area of applicability of the model \( M \) in the data subset
\( j \); \( E^{M,low}_{j} \) is the lowest value of the energy in
the area of applicability of the model \( M \) in the data subset
\( j \), and \( w_{j} \) is the weight determined from the best
fit in the same interval (hence \( w_{j} \) will depend itself on
\( E^{M,high}_{j} \) and \( E^{M,low}_{j} \)). 
In our case the applicability
indicator takes the form: \begin{eqnarray}
A^{M}= & {1\over 9} & \left( A^{M}_{pp,\sigma }+A^{M}_{{\overline{p}}p,\sigma
}+A^{M}_{\pi ^{+}p,\sigma }+A^{M}_{\pi ^{-}p,\sigma }+A^{M}_{K^{+}p,\sigma
}+A^{M}_{K^{-}p,\sigma }+\right. \nonumber \\
 &  & \nonumber \\
 & + & \left. A^{M}_{\Sigma ^{-}p,\sigma }+A^{M}_{\gamma p,\sigma
}+A^{M}_{\gamma \gamma ,\sigma }\right) \label{AM}.
\end{eqnarray}
Inspection of the fit results shows that for some modification of
the parametrisations we obtain rather good fits starting from $E_{min} = 4$
or $5 \ {\rm GeV}$ but with negative contributions to the total cross
sections from terms corresponding to the exchange of the pomeron-like
objects at low energy part of the area of applicability as defined
above. This is unphysical and we are forced to add an additional constraint
to the area of applicability: We exclude from it the low energy part
where at least in one collision there is a negative contribution from
the total sum of the pomeron-like (asymptotically rising) terms. The
situation is illustrated in Tables A1.3 and A2.3 of the appendices
where excluded intervals are marked by minus as upper case index at
the $\chi^2/dof$ value. It is interesting that some models turned
out to have an empty area of applicability once this criterion was
imposed.
\item [(2)~Confidence-1~Indicator.]\[
C^{M}_{1}=CL\%\]
 where the CL refers to the whole area of applicability of the model
M.
\item [(3)~Confidence-2~Indicator.]\[
C^{M}_{2}=CL\%\]
 where the CL refers to the intersection of the areas of applicability
of all models qualified for the comparison (we choose here \( \sqrt{s}\geq 5 \)
GeV for the fits without \( \rho  \) parameter (see Table A1.3) and
\( \sqrt{s}\geq 9 \) GeV for the fits with \( \rho  \) data (see
Table A2.3).
\item [(4)~Uniformity~Indicator.]This indicator measures the variation
of the \( \chi ^{2}/nop \) from bin to bin for some data binning
motivated by physics: \begin{equation}
\label{U}
U^{M}=\left\{ {1\over N_{sets}}\sum _{j}{1\over 4}\left[ \frac{\chi
_{R}^{2}(t)}{N^{t}_{nop}}-\frac{\chi _{R}^{2}(j)}{N^{j}_{nop}}\right]
^{2}\right\} ^{-1},
\end{equation}
 where \( t \) denotes the total area of applicability, \( j \)
is a multi-index  denoting the pair (data set, observable).
In our case we use the calculation of the \( \chi _{R}^{2}/nop \)
for each collision separately, i.e. the sum runs as in the case of
the applicability indicator.
\item [(5)~Rigidity~Indicator.]As the measure of the rigidity of the model
we propose to use the indicator \begin{equation}
\label{R1}
R^{M}_{1}={N_{dp}^{M}(A)\over {1+N^{M}_{par}}}
\end{equation}
 The most rigid model has the highest value of the number of data
points per adjustable parameter. The exact theory \( T \) (with no
adjustable parameters) has the rigidity value \( R^{T}=N_{dp}^{M}(A) \)
--- the total number of data points in the area of applicability. This
indicator takes into account the set of known regularities in the
data that were incorporated into the model to reduce the number of
adjustable parameters and to increase the statistical reliability
of the parameter estimates.
\item [(6)~Reliability~Indicator.]\begin{equation}
\label{R2}
R^{M}_{2}={2\over N_{par}(N_{par}-1)}\cdot \sum _{i>j=1}^{N}\Theta
(90.0-C^{R}_{ij})
\end{equation}
 where \( C^{R}_{ij} \) -- is the correlation matrix element in \( \% \)
calculated in the fit at the low edge of the applicability area. This
indicator characterizes the goodness of the parameter error matrix.
For the diagonal correlator this indicator is maximal and equals \( 1 \).
\item [(7)~Stability-1~Indicator.]\begin{equation}
\label{S1}
S^{M}_{1}=\left\{ {1\over N_{steps}N_{par}^{M}}\sum _{steps}\sum
_{ij}(P^{t}-P^{step})_{i}(W^{t}+W^{step})^{-1}_{ij}(P^{t}-P^{step})_{j}\right\}
^{-1}
\end{equation}
 where: \( P^{t} \) - vector of parameters values obtained from the
model fit to the whole area of applicability;

\( P^{step} \) - vector of parameters values obtained from the model
fit to the reduced data set on the \( step \), in our case \( step \)
means shift in the low edge of the fit interval to the right by 1
GeV, if there are no steps then \( S^{M}_{1}=0 \) by definition;

\( W^{t} \) and \( W^{step} \) are the error matrix estimates obtained
from the fits to the total and to the reduced on the step \( s \)
data samples from the domain of applicability.

\end{description}
We give the results of these comparisons in Table 2 and Appendix 1, Table A1.2.

The development of these indicators is needed to allow us to verify
automatically the rough features of a large quantity of models (see
Appendices 1 and 2). Hence, as a first {}``numerical trigger{}''
to indicate the best fits, we have adopted a simple ranking scheme,
which complements the usual {}``best \( \chi ^{2} \){}'' criterion.
As all the features measured by the indicators are highly desirable,
we adopt for the rank, in a given ensemble of models, a definition
that gives equal weight to all indicators
\begin{equation}
I^{m}_{k}=(A^{m},C^{m}_{1},C^{m}_{2},U^{m},R^{m}_{1},
R^{m}_{2},S^{m}_{1})\label{IK}
\end{equation}
 where the index \( m \) describes the model, index \( k \) describes
the indicator type.

Having calculated all components of the indicators, it is easy, for
a given indicator, to assign a number of points to a given model \( M \):
\begin{equation}
\label{PMK}
P^{M}_{k}=\sum _{m\neq M}{(2\Theta (I^{M}_{k}-I^{m}_{k})+\delta
_{I^{M}_{k},I^{m}_{k}})},
\end{equation}
The rank of models is then obtained via the total amount of points
of the model : \begin{equation}
\label{PM}
P^{M}=\sum _{k}{P^{M}_{k}}=\sum _{k}\sum _{m\neq M}{(2\Theta
(I^{M}_{k}-I^{m}_{k})+\delta _{I^{M}_{k},I^{m}_{k}})}
\end{equation}
 In this approach, the best models are the models with the highest
\( P^{M} \) value. In the Tables 2 and 5, and in the appendices,
we present the ranking of 33  recently discussed parametrisations:
28 of them had a sufficiently high CL for comparison on the \( \sigma _{tot}
\)-data
and 21 of them had a sufficiently high CL for comparison on the \( \sigma
_{tot}(s) \)-
and \( \rho (s) \)-data.

\bigskip{}

{\centering {\small \begin{tabular}{|l||c|c|c|c|c|c|c|c|}
\hline
{\small Model Code}&
{\small \( A^{M} \)}&
{\small \( C^{M}_{1} \)}&
{\small \( C^{M}_{2} \)}&
{\small \( U^{M} \)}&
{\small \( R^{M}_{1} \)}&
{\small \( R^{M}_{2} \)}&
{\small \( S^{M}_{1} \)}&
\multicolumn{1}{c|}{rank \( P^{M} \)}\\
\hline
\hline
\multicolumn{9}{|c|}{{\small simple pole: }}\\
\hline
{\small \( {\textrm{RR E}_{nf}(19)} \)}&
\textbf{\small 2.6}&
{\small 91. }&
{\small 81. }&
{\small 51. }&
{\small 25. }&
{\small 0.88 }&
\textbf{\small 0.18} {\small }&
208\\
\hline
{\small \( {\textrm{RR E}^{qc}(17)} \)}&
\textbf{\small 2.6}&
{\small 86.}&
{\small 79. }&
{\small 88. }&
{\small 28. }&
\textbf{\small 0.94}&
{\small 0.15 }&
\textbf{252}\\
\hline
\hline
\multicolumn{9}{|c|}{{\small simple+double pole }}\\
\hline
{\small \( {\textrm{RRL}_{nf}(19)} \)}&
\textbf{\small 2.6}&
{\small 76. }&
{\small 95. }&
{\small 36. }&
\textbf{\small 29.}&
{\small 0.79 }&
\textbf{\small 0.16 }&
212\\
\hline
{\small \( {\textrm{RRPL}(21)} \)}&
{\small 2.2}&
{\small 65. }&
{\small 99.7 }&
{\small 59. }&
{\small 26. }&
{\small 0.81 }&
{\small 0.082 }&
162\\
\hline
\hline
\multicolumn{9}{|c|}{{\small simple+triple pole }}\\
\hline
{\small \( {(\textrm{RR})^{d}\textrm{ P$_{nf}$L}2(20)} \)}&
{\small 2.5}&
{\small 59. }&
\textbf{\small 99.9} {\small }&
{\small 38. }&
{\small 28. }&
{\small 0.88 }&
{\small 0.098}&
120\\
\hline
{\small \( {\textrm{RRP$_{nf}$L}2_{u}(21)} \)}&
{\small 2.5}&
{\small 68. }&
{\small 99.7 }&
{\small 34. }&
{\small 26. }&
{\small 0.91 }&
{\small 0.008}&
182\\
\hline
{\small \( {(\textrm{RR})^{d}\textrm{ PL}2_{u}(17)} \)}&
\textbf{\small 2.6}&
\textbf{\small 99.8}&
{\small 99.7}&
\textbf{\small 185.}&
{\small 28. }&
{\small 0.88 }&
\textbf{\small 0.16} {\small }&
\textbf{296}\\
\hline
\end{tabular}}\small \par}

\begin{quote}
\medskip{}
{\small Table 2: model quality indicators for the models kept in Table
1. Bold-faced characters indicate the best model for a given indicator.}{\small
\par}\bigskip{}

\end{quote}

On the other hand, it is also possible to use these indicators directly,
as characterizing each model. For instance, if we analyse the first
two lines of Table 2, we directly see from column 1 that simple-pole
models apply in as big an energy band as the other models. The second
and third columns tell us however that the best CL are achieved by
triple-pole models with the double-pole models closely behind. The
fourth column tells us that while most models do not reproduce all
data equally well (see also Table 4), the most uniform model is 
(RR)$_d$PL2$_u$(17).
The fifth column indicates that the models apply in similar energy
ranges and have similar numbers of parameters. Similarly, we see from
the sixth column that the reliability of the error matrices is similar.
However, the seventh column clearly indicates that the parameters
of RRPL(21), {\small \( {(\textrm{RR})^{d}\textrm{ P$_{nf}$L}2(20)} \)}
and {\small \( {\textrm{RRP$_{nf}$L}2_{u}(21)} \)} are very sensitive to
the minimum energy considered, and hence that these models are not
stable w.r.t. that minimum energy.

\section{Fits to all lower-energy forward data}

Given that the fits to total cross sections are unable to decide on
the singularity structure of the amplitudes, one can turn to other
data, namely the real part of the forward amplitude. It can be obtained
through analyticity and $s\rightarrow$ crossing symmetry 
from the form of the cross section (see Appendix 3). If one keeps
the same minimum energy, then a joint fit to both cross sections and
real parts reaches a very different conclusion. We show in Table 3
the models which achieve a \( \chi ^{2}/dof \) less than 1 for \( \sqrt{s}\geq
5 \)
GeV.\bigskip{}

{\centering {\small \begin{tabular}{|l|c|c|c|c|c|c|c|c|}
\cline{2-9}
\multicolumn{1}{l|}{}&
\multicolumn{8}{c|}{ {\small \( \sqrt{s_{min}} \) in GeV and number of data
points }}\\
\hline
{\small \( {\textrm{Model code}\, (\textrm{N}_{par})} \)}&
{\small 3 (904)}&
{\small 4 (742)}&
{\small 5 (648)}&
{\small 6 (569) }&
{\small 7 (498) }&
{\small 8 (453) }&
\multicolumn{1}{c|}{{\small 9 (397) }}&
{\small 10 (329) }\\
\hline
\hline
{\small \( \textrm{RRE}_{nf} (19)\)}&
{\small 1.8}&
{\small 1.4 }&
{\small 1.1 }&
{\small 1.1 }&
{\small 1.1 }&
{\small 1.1 }&
{\small 1.0 }&
{\small 1.0}\\
\hline
\hline
{\small \( {\textrm{RRL}_{nf}(19)} \)}&
{\small 1.6 }&
{\small 1.1 }&
{\small \( \bf 0.97 \)}&
\textbf{\small \( \bf 0.97 \)}&
\textbf{\small 1.0} {\small }&
{\small \( \bf 0.96 \)}&
{\small \( \bf 0.94 \)}&
{\small \( \bf 0.93 \)}\\
\hline
{\small \( {\textrm{RRPL}}(21) \)}&
{\small 1.6 }&
{\small 1.1 }&
{\small \( \bf 0.98 \)}&
{\small \( \bf 0.98 \)}&
{\small \( \bf 0.99 \)}&
{\small \( \bf 0.94 \)}&
{\small \( \bf 0.93 \)}&
{\small \( \bf 0.91 \)}\\
\hline
\hline
{\small \( {(\textrm{RR})^{d}\textrm{ P$_{nf}$ L}2}(20) \)}&
{\small 1.9 }&
{\small 1.2 }&
{\small \( \bf 1.0 \)}&
{\small \( \bf 1.0 \)}&
{\small \( \bf 0.99 \)}&
{\small \( \bf 0.94 \)}&
{\small \( \bf 0.93 \)}&
{\small \( \bf 0.92 \)}\\
\hline
{\small \( {\textrm{RRP$_{nf}$L}2_{u}}(21) \)}&
{\small 1.8 }&
{\small 1.1 }&
{\small \( \bf 0.97 \)}&
{\small \( \bf 0.97 \)}&
{\small \( \bf 0.97 \)}&
{\small \( \bf 0.92 \)}&
{\small \( \bf 0.93 \)}&
{\small \( \bf 0.92 \)}\\
\hline
{\small \( {(\textrm{RR})^{d}\textrm{ PL}2_{u}}(17) \)}&
{\small 2.0 }&
{\small 1.3 }&
{\small 1.0 }&
{\small \( \bf 1.0 \)}&
{\small \( \bf 0.98 \)}&
{\small \( \bf 0.94 \)}&
{\small \( \bf 0.93 \)}&
\textbf{\small \( \bf 0.93 \) }\\
\hline
\end{tabular}}\small \par}
\begin{quote}
\medskip{} {\small Table 3: representative models fitting all cross
section and \( \rho  \) data down to 5 GeV. Numbers in bold represent
the area of applicability of each model.} \bigskip{}
\end{quote}

The clearest outcome of this is that all models with a simple pole
pomeron are then eliminated. The best \( \chi ^{2}/dof \) for these
is 1.12 for RRE\( _{nf} \). Although these values may not seem too
problematic, one has to realise that we are fitting to a large number
of data points (648 for \( \sqrt{s}>5 \) GeV), hence this model is
rejected at the 98\% C.L.

\subsection{Evaluation of the dataset}

However, one needs to check where these values of \( \chi ^{2}/dof \)
come from. Hence we can look in detail at the various processes and
quantities fitted to. We show in Table 4 the results of 3 representative
models. The first two are kept in Table 3, whereas we came to the
conclusion that the third is excluded. We see that the main difference
comes from the \( \rho  \) parameter data, which are much better
fitted by the first two models than by the third. However, it is rather
difficult to reach a definite conclusion, given the fact that these
data are not perfectly fitted by any model: in particular, the \( \pi p \)
and \( pp \) data. \bigskip{}

{\centering \begin{tabular}{|c|c|c|c|c|}
\hline
Reaction &
Number of &
RRP$_{nf}$L2$_u$ &
RRPL &
RRE$_{nf}$ \\
&
data points&
&
&
\\
\hline
\hline
\( \sigma _{pp} \)&
 112 &
0.87&
0.87&
0.89 \\
\hline
\( \sigma _{\overline{p}p} \)&
 59 &
1.2&
1.0 &
1.1\\
\hline
\( \sigma _{\pi ^{+}p} \)&
 50 &
0.78&
0.78 &
1.4\\
\hline
\( \sigma _{\pi ^{-}p} \)&
 106 &
0.89 &
0.90 &
0.88 \\
\hline
\( \sigma _{K^{+}p} \)&
 40 &
0.71 &
0.72 &
1.0 \\
\hline
\( \sigma _{K^{-}p} \)&
 63 &
0.61 &
0.62 &
0.72 \\
\hline
\( \sigma _{\Sigma ^{-}p} \)&
 9 &
0.38 &
0.38 &
0.39 \\
\hline
\( \sigma _{\gamma p} \)&
 38 &
0.62 &
0.75 &
0.59\\
\hline
\( \sigma _{\gamma \gamma } \)&
 30 &
0.7 &
0.95 &
0.55 \\
\hline
\hline
\( \rho _{pp} \)&
 74 &
1.8&
1.6&
1.8\\
\hline
\( \rho _{\overline{p}p} \)&
 11 &
0.55 &
0.47 &
0.60 \\
\hline
\( \rho _{\pi ^{+}p} \)&
 8 &
1.5&
1.6&
2.7\\
\hline
\( \rho _{\pi ^{-}p} \)&
 30 &
1.2&
1.3&
2.1\\
\hline
\( \rho _{K^{+}p} \)&
 10 &
1.0&
1.1&
0.83 \\
\hline
\( \rho _{K^{-}p} \)&
 8  &
0.96 &
1.2&
1.8 \\
\hline
\end{tabular}\par}

\begin{quote}
\medskip{}{\small Table 4: The values of the \( \chi ^{2} \) per
data point (\( \chi ^{2}/nop \)) for each process in three representative
models, for \( \sqrt{s}>5 \) GeV.}\bigskip{}
\end{quote}

\subsection{Best models for all forward data}

We can generalize the previous quality indicators to the full set
of forward data. We give in Table 5 and in Appendix 2 the quality
indicators for representative models fitting both total cross sections
and \( \rho  \) parameters. We have introduced a second stability
indicator, \( S_{2} \), which is analogous
to the stability-1 indicator \begin{equation}
\label{S2}
S^{M}_{2}=\left\{ {1\over 2N_{par}^{M}}\sum _{ij}(P^{t}-P^{t(no\: \rho
)})_{i}(W^{t}+W^{t(no\: \rho )})^{-1}_{ij}(P^{t}-P^{t(no\: \rho )})_{j}\right\}
^{-1}.
\end{equation}
 In this case, we fit the whole set of the model parameters to the
full area of applicability (superscript $t$) and the same set of
parameters but to the data sample without $\rho$-data (superscript
$t(no\ \rho)$).~ This indicator characterizes the reproducibility
of the parameters values when fitting to the reduced data sample and
reduced number of observables but with the same number of adjustable
parameters. This indicator might be strongly correlated with the uniformity
indicators. We add $S_2^m$ to the list of indicators entering $I^{m}_{k}$ in 
Eq.~(\ref{IK}) when 
we determine the best models for the full set of data, and run the sums for
all indicators for 15 sets of data instead of 9, as we now include the real 
parts of $pp$, $\bar p p$, $K^\pm p$ and $\pi^\pm p$.

As we can see, the two parametrisations based on double poles and
on triple poles achieve comparable levels of quality, and one cannot
decide which is the best based on these indicators. In the conclusion,
we shall explain which physics arguments lead us to prefer the triple
pole alternative.
\bigskip{}

{\centering {\small \begin{tabular}{|l||c|c|c|c|c|c|c|c|c|}
\hline
{\small Model Code}&
{\small \( A^{M} \)}&
{\small \( C^{M}_{1} \)}&
{\small \( C^{M}_{2} \)}&
{\small \( U^{M} \)}&
{\small \( R^{M}_{1} \)}&
{\small \( R^{M}_{2} \)}&
{\small \( S^{M}_{1} \)}&
{\small \( S^{M}_{2} \)}&
rank  \( P^{M} \)\\
\hline
\hline
{\small \( {\textrm{RRP$_{nf}$L}2_{u}(21)} \)}&
\textbf{2.2}&
{\small 68. }&
\textbf{85.} {\small }&
\textbf{23.} {\small }&
{\small 29. }&
\textbf{0.90}&
{\small 0.22 }&
{\small 0.10 }&
\textbf{222}\\
\hline
{\small \( {(\textrm{RR})^{d}\textrm{ P$_{nf}$L}2(20)} \)}&
\textbf{2.2}&
{\small 50. }&
{\small 82. }&
{\small 18. }&
{\small 31. }&
\textbf{0.90} {\small }&
{\small 0.27 }&
{\small 0.41 }&
178\\
\hline
{\small \( {(\textrm{RR})^{d}\textrm{ PL}2_{u}}(17) \)}&
2.0&
{\small 50. }&
\textbf{83.} {\small }&
{\small 16. }&
\textbf{32.} {\small }&
{\small 0.88 }&
\textbf{0.30} {\small }&
{\small 0.67 }&
174\\
\hline
\hline
{\small \( {\textrm{RRL}_{nf}(19)} \)}&
{\small 1.8}&
\textbf{73.}&
{\small 81. }&
{\small 17. }&
\textbf{32.} {\small }&
{\small 0.78 }&
\textbf{0.29 }&
\textbf{1.3 }&
\textbf{222}\\
\hline
{\small \( {\textrm{RRPL}(21)} \)}&
{\small 1.6}&
\textbf{67.} {\small }&
{\small 82. }&
\textbf{26.} {\small }&
{\small 29. }&
{\small 0.75 }&
{\small 0.21 }&
\textbf{1.1  }&
173\\
\hline
\end{tabular}}\small \par}

{\centering \medskip{} {\small Table 5: Quality indicators in five
representative models fitting well all forward data.}\small \par}

{\centering \bigskip{}\par}

\section{Other models}

We have tried to impose the Johnson-Treiman-Freund \cite{jt, freund}
relation for the cross section differences \( \Delta \sigma (N)=5\Delta \sigma
({\pi }),\Delta \sigma (K)=2\Delta \sigma ({\pi }) \),
and the models corresponding to this are marked by an index \( c \)
in Appendices 1 and 2. These rules, while not being totally excluded,
never lead to an improvement of the fit, and in some case degrade
the fit considerably. It is interesting to note however that they
produce the two parametrisations with fewest parameters acceptable
above 8 GeV.

We also considered alternative models which have been proposed or
rediscovered recently \cite{Feigenbaum:1997gy, Lipkin:1999tc}, and
confront them with our full dataset. From Table 6, one sees clearly
that the parameter values and possibly the model themselves have practically
zero confidence levels at all starting collision energies \( \sqrt{s_{min}} \)
from 3 to 10 GeV.

\bigskip{}
{\centering \begin{tabular}{|l|c|c|c|c|c|c|c|}
\cline{2-8}
\multicolumn{1}{l|}{}&
\multicolumn{7}{c|}{ \( \sqrt{s_{min}} \)~~~~in GeV }\\
\hline
\( {\textrm{Model code}\, (\textrm{N}_{par})} \)&
3 &
4 &
5 &
6 &
7 &
8 &
9 \\
\hline
\hline
FFP-97\cite{Feigenbaum:1997gy}&
101 &
16.26 &
3.28 &
2.3 &
2.3 &
2.39 &
2.34 \\
\hline
Lipkin~TCP\cite{Lipkin:1999tc}&
4.63 &
3.14 &
2.54 &
2.61 &
2.86 &
3.07 &
3.48  \\
\hline
\end{tabular}\par}

\medskip{}
{\centering {\small Table 6: \( \chi ^{2}/dof \) of two excluded
parametrisations.}\small \par}
\bigskip{}

\section{Other data}

As in the previous studies \cite{Cudell:2000tx} of fitting the data
sample \cite{Groom:2000in}, we have also excluded all cosmic data
points \cite{Baltrusaitis:1984ka}, \cite{Honda:1993kv} in this study
of the analytic amplitude models. There are two reasons for that:
the original numerical Akeno (Agasa) data are not available and there
are the contradictory statements \cite{Durand:1988cr, Kopeliovich:1989iy,
Nikolaev:1993mc, Velasco:1999ce, Block:2000pg}concerning
the cross section values of the cosmic data points from both Fly's
Eye and Akeno(Agasa).

Having selected the models which reproduce best the accelerator data,
we are now able to clarify how well they meet the three cosmic rays
data samples. For each cosmic data samples, i.e. those of the original
experiments \cite{Baltrusaitis:1984ka}, \cite{Honda:1993kv}; those
corrected by Nikolaev et al. \cite{Kopeliovich:1989iy}, \cite{Nikolaev:1993mc};
and those corrected by Block et al. \cite{Block:2000pg} (see also
\cite{Durand:1988cr}), we calculate the \( \chi ^{2}/nop \) for each
model with parameters fixed at the beginning of their areas of applicability
defined by accelerator data. The results are shown in Table 7.

It turns out that the original cosmic experimental data are best fitted
by our high-rank parametrisations. The data sample corrected by Block
et al. data is also fitted well, as the data points were lowered within
the limits of the uncertainties reported in the original experimental
publications.

\bigskip{}

\bigskip{}
{\centering \begin{tabular}{|c|r|c|r|c|r|c|}
\cline{2-3} \cline{4-5} \cline{6-7}
\multicolumn{1}{c|}{}&
\multicolumn{2}{|c|}{ Experiment }&
\multicolumn{2}{|c|}{ Nikolaev et al. }&
\multicolumn{2}{|c|}{ Block et al. }\\
\hline
Model Code &
\( \chi ^{2} \)&
\( \chi ^{2}/nop \)&
\( \chi ^{2} \)&
\( \chi ^{2}/nop \)&
\( \chi ^{2} \)&
\( \chi ^{2}/nop \)\\
\hline
\hline
\( {\textrm{RRP$_{nf}$L}2_{u}(21)} \)&
1.62 &
\( 0.23 \)&
14.31 &
2.04 &
3.30 &
\( 0.47 \)\\
\hline
\( {(\textrm{RR})^{d}\textrm{P$_{nf}$L}2_{u}(19)} \)&
1.73 &
\( 0.25 \)&
13.96 &
1.99 &
3.45 &
\( 0.49 \)\\
\hline
\hline
\( {\textrm{RRL}_{nf}(19)} \)&
2.52 &
\( 0.36 \)&
24.25 &
3.46 &
2.19 &
\( 0.31 \)\\
\hline
\( {\textrm{RRPL}(21)} \)&
2.93 &
\( 0.42 \)&
25.48 &
3.64 &
2.34 &
\( 0.33 \) \\
\hline
\end{tabular}\par}

\begin{quote}
\medskip{}
{\small Table 7: the \( \chi ^{2} \) of the cosmic ray data, corrected
in several different ways \cite{Durand:1988cr, Kopeliovich:1989iy,
Nikolaev:1993mc, Velasco:1999ce, Block:2000pg},
for each of the best parametrisations fitting the accelerator data.}{\small
\par}\bigskip{}

\end{quote}
\section{Analysis and conclusion}

The above analysis shows that there are several scenarios which can
account for the observed forward hadronic scattering amplitudes. These
scenarios all have their merits, and some of them have problems. Although
only preliminary conclusions can be drawn based on these data, we
can outline these various possibilities, and present their consequences.

\subsection{Possible parametrisations}

The three possible scenarios consist of simple, double or triple poles
in the complex \( J \) plane accounting for the rising part of the
cross section. We give in Table 8 the parameters of each model. All
have the same parametrisation for the exchange of the leading meson
trajectories, but the values of the various intercepts and residues
are very different. The \( C=-1 \) part of the amplitude is rather
stable, but the \( C=+1 \) part turns out to be very model-dependent
as it mixes with the pomeron contribution, with in some cases much
larger values of the intercept \( \alpha _{1} \) than those normally
expected from duality-breaking in strong interaction physics. Because
of this, the lower energy data cannot fix the nature of the pomeron
as the details of the \( a/f \) contribution are not known. The data
for \( \Sigma p \) scattering sometimes lead to a negative $a/f$
contribution, which is incompatible with Regge theory, and to an extrapolation
at high energy that overshoots the \( pp \) and \( \overline{p}p \)
cross sections. However, the size of the error bars clearly shows
that acceptable values are allowed and that these data do not introduce
much of a constraint on the fit.

\subsubsection{Simple poles}

The first scenario is the simplest conceptually: the pomeron would
correspond to some glueball trajectory, and have properties similar
to those of the mesons. This model has the advantage that it must
then factorize, and hence it can be generalized easily and successfully
to many other processes. The residues of the pomeron can also be made
totally compatible with quark counting.

It provides good fits to all data for \( \sqrt{s}\geq 9 \) GeV, acceptable
fits for the total cross sections for \( \sqrt{s}\geq 5 \) GeV, but
fails to reproduce both the total cross section and the \( \rho  \)
parameter for \( \sqrt{s}\geq 5 \) GeV. One can of course take the
attitude that the data have problems, and not include them, or that
there are sub-dominant effects at these energies, and that it is natural
for the model not to be extended so low. On the other side of the
energy spectrum, one expects to have unitarity corrections at very
large energies. In practice, however, this model differs by a few
percents from the RRPL2\( _{u} \) parametrisation, mentioned below,
up to LHC energies, and hence unitarizing corrections do not need
to be introduced yet.

This model shows a non-degeneracy of the dominant meson trajectories,
with somewhat larger \( a/f \) intercept \( \alpha _{1} \) and somewhat
smaller \( \rho /\omega  \) intercept \( \alpha _{2} \), which may
well be compatible with the known trajectories.

Furthermore, it is well known that one needs to introduce a new simple
pole to account for DIS data in such a scenario. Such a new rising
term seems to be totally absent from the soft data, which seems rather
odd, but cannot be ruled out. We give in Table 8, column 3, the best
parameters for this model in the fit to total cross sections.
\bigskip{}

\subsubsection{Double poles}

One can also assume that the amplitude contains a double pole at \( J=1 \).
This then provides for a rising \( \ln s \) term in the total cross
section, as well as a constant term. This kind of parametrisation
(shown in Table 8, column 2) gives excellent fits to the soft data,
and can be extended to deep-inelastic scattering \cite{Desgrolard}
without any further singularity. Furthermore, it never violates unitarity,
and hence it can be extended to arbitrarily large energies.

However, it suffers from several drawbacks. First of all, the pomeron
term becomes negative below 9.5 GeV, and hence processes which couple
only to the pomeron by Zweig's rule would have negative cross sections
if one uses factorization. However, the latter is proven only for
simple poles, and hence this problem is not a sufficient reason to
reject these parametrisations. Similarly, the split of the leading
meson trajectories is quite big, somewhat bigger than what a normal
duality-breaking estimate or a linear extrapolation of the known resonances
would allow \cite{Desgrolard:2001sf}. As a result, the pomeron in
this class of variants is inevitably compromising with the crossing
even reggeon in the Regge region in the sense that it must effectively
counter-balance the excessive contribution of the reggeon. Thus the
pomeron term in this case may be representing more than the asymptotic
behaviour of the amplitude. One may therefore say that a pomeron associated
with reasonably degenerate reggeons may be more natural from the point
of view of duality. But again, one cannot prove linearity of the trajectories,
hence the model may survive. Finally, it seems that quark counting
is respected to a very good approximation by the coefficients of the
log and of the constant term. This only reinforces the problem of
negativity as it is very difficult to conceive a non factorizing pole
which would nevertheless respect quark counting.

\subsubsection{Triple poles}

Finally, the best fits are given by models that contain a triple pole
at \( J=1 \), which then produce \( \ln ^{2}s \), \( \ln s \)
and constant terms in the total cross section. The best parameter
values for this model are given in Table 8, column 1. The most interesting
properties may be that the constant term respect quark counting to
a good approximation, whereas the ln\( ^{2}s \) term can be taken
as universal, i.e. independent of the process, as advocated in
\cite{solovev:1973, Ter-Martirosian:1988yy}
and rediscovered in \cite{Gauron:2000ri} (see also \cite{log2 very high E}).
The universality of the rising term is expected in the case of the
eikonal unitarisation of a bare pomeron with the intercept larger
than 1, because the coefficient of the rising term turns out to depend
only on the intercept and slope of the bare pomeron \cite{FFKT}.
But for the J-plane singularities of double and triple pole types
considered in this paper, the structure of such a singularity \cite{Gauron 88}
and the origin of its universality is less obvious. Nevertheless,
such a singularity at \( J=1 \) may in fact have a theoretical explanation:
recently, Bartels, Lipatov and Vacca \cite{Bartels:2000yt} discovered
that there are, in fact, two types of Pomeron in LLA : besides the
well-known BFKL pomeron associated with 2-gluon exchanges, and with
an intercept bigger than 1, there is a second one associated with
\( C=+1 \) three-gluon exchanges and having an intercept precisely
located at 1. It is tempting to speculate that, after unitarisation
is performed in the gluon sector, the BFKL pomeron would finally lead
to a universal Heisenberg-type pomeron, exclusively connected with
the gluon sector.

Furthermore, the degeneracy of the lower trajectories is respected
to a very good approximation, and the model seems extendible to deep
inelastic scattering \cite{Cudell 2001}. This model also respect
unitarity by construction.

One must note that in some processes, the falling $\ln^2(s/s_0)$ term from the 
triple pole at $s<s_0$ is important in restoring the degeneracy of 
the lower trajectories
at low energy. Hence the squared logarithm manifests itself not only
at very high energies, but also at energies below its zero.

Hence we feel that this solution is the one that currently meets all
phenomenological and theoretical requirements.

\bigskip{}
{\centering \begin{tabular}{|c|c|c||c|c|c||c|c|c|}
\hline
\multicolumn{1}{|c||}{ {\footnotesize Model }}&
\multicolumn{2}{|c||}{ {\footnotesize RRP$_{nf}$L2\( _{u} \) }}&
\multicolumn{3}{|c||}{ {\footnotesize RRL\( _{nf} \)} }&
\multicolumn{3}{|c|}{ {\footnotesize RRE\( _{nf} \) }}\\
\hline
\multicolumn{1}{|c||}{{\footnotesize \( \chi ^{2}/dof \)}}&
\multicolumn{2}{|c||}{{\footnotesize \( 0.97 \)}}&
\multicolumn{3}{|c||}{{\footnotesize 0.97} }&
\multicolumn{3}{|c|}{{\footnotesize 1.12}}\\
\hline
\multicolumn{1}{|c||}{{\footnotesize CL{[}\%{]}}}&
\multicolumn{2}{|c||}{{\footnotesize 67.98 }}&
\multicolumn{3}{|c||}{{\footnotesize 73.37} }&
\multicolumn{3}{|c|}{{\footnotesize 2.08 }}\\
\hline
\cline{1-1}
\multicolumn{1}{|c|}{{\footnotesize Parameter} }&
{\footnotesize Mean }&
{\footnotesize Uncertainty }&
{\footnotesize Param.}&
{\footnotesize Mean }&
{\footnotesize Uncert.} &
{\footnotesize Param. }&
{\footnotesize Mean} &
{\footnotesize Uncert.} \\
\hline
{\footnotesize \( s_{0} \)}&
{\footnotesize 34.0 }&
{\footnotesize 5.4 }&
{\footnotesize \( A \)}&
{\footnotesize -30.3} &
{\footnotesize 3.6} &
{\footnotesize \( \alpha _{\wp } \)}&
{\footnotesize 1.0959} &
{\footnotesize 0.0021} \\
{\footnotesize \( \alpha _{1} \)}&
{\footnotesize 0.533} &
{\footnotesize 0.015 }&
{\footnotesize \( \alpha _{1} \)}&
{\footnotesize 0.7912 }&
{\footnotesize 0.0080 }&
{\footnotesize \( \alpha _{1} \)}&
{\footnotesize 0.6354} &
{\footnotesize 0.0095} \\
{\footnotesize \( \alpha _{2} \)}&
{\footnotesize 0.4602} &
{\footnotesize 0.0064} &
{\footnotesize \( \alpha _{2} \)}&
{\footnotesize 0.4555 }&
{\footnotesize 0.0063} &
{\footnotesize \( \alpha _{2} \)}&
{\footnotesize 0.4420 }&
{\footnotesize 0.0099 }\\
{\footnotesize \( Z^{pp} \)}&
{\footnotesize 35.83} &
{\footnotesize 0.40 }&
{\footnotesize \( B \)}&
{\footnotesize 6.71 }&
{\footnotesize 0.22 }&
{\footnotesize \( X^{pp} \)}&
{\footnotesize 18.45} &
{\footnotesize 0.41 }\\
{\footnotesize \( Z^{\pi p} \)}&
{\footnotesize 21.23} &
{\footnotesize 0.33 }&
{\footnotesize \( \lambda _{\pi p} \)}&
{\footnotesize 0.6833 }&
{\footnotesize 0.0045} &
{\footnotesize \( X^{\pi p} \)}&
{\footnotesize 11.74 }&
{\footnotesize 0.24 }\\
{\footnotesize \( Z^{Kp} \)}&
{\footnotesize 18.23 }&
{\footnotesize 0.30 }&
{\footnotesize \( \lambda _{Kp} \)}&
{\footnotesize 0.6429} &
{\footnotesize 0.0073} &
{\footnotesize \( X^{Kp} \)}&
{\footnotesize 10.45} &
{\footnotesize 0.19} \\
{\footnotesize \( Z^{\Sigma p} \)}&
{\footnotesize 35.6 }&
{\footnotesize 1.4} &
{\footnotesize \( \lambda _{\Sigma p} \)}&
{\footnotesize 1.059} &
{\footnotesize 0.056} &
{\footnotesize \( X^{\Sigma p} \)}&
{\footnotesize 18.44} &
{\footnotesize 1.1 }\\
{\footnotesize \( Z^{\gamma p} \)}&
{\footnotesize 29.4} &
{\footnotesize 3.0} &
{\footnotesize \( \lambda _{\gamma p} \)}&
{\footnotesize 0.00356 }&
{\footnotesize 0.000048 }&
{\footnotesize \( X^{\gamma p} \)}&
{\footnotesize 0.0592 }&
{\footnotesize 0.0012} \\
{\footnotesize \( Z^{\gamma \gamma } \)}&
{\footnotesize 20.4} &
{\footnotesize 5.0 }&
{\footnotesize \( \lambda _{\gamma \gamma } \)}&
{\footnotesize 9.37 10\( ^{-6} \)}&
{\footnotesize 5.2 10\( ^{-7} \)}&
{\footnotesize \( X^{\gamma \gamma } \)}&
 {\footnotesize 0.0001619}&
{\footnotesize 9.7 10\( ^{-6} \)}\\
{\footnotesize \( Y^{pp}_{1} \)}&
{\footnotesize 42.1 }&
{\footnotesize 1.3 }&
{\footnotesize \( Y^{pp}_{1} \)}&
{\footnotesize 105.8} &
{\footnotesize 2.9} &
{\footnotesize \( Y^{pp}_{1} \)}&
{\footnotesize 66.1 }&
{\footnotesize 1.2 }\\
{\footnotesize \( Y^{pp}_{2} \)}&
{\footnotesize 32.19} &
{\footnotesize 0.94 }&
{\footnotesize \( Y^{pp}_{2} \)}&
{\footnotesize 33.36} &
{\footnotesize 0.96 }&
{\footnotesize \( Y^{pp}_{2} \)}&
{\footnotesize 35.3 }&
{\footnotesize 1.6} \\
{\footnotesize \( Y^{\pi p}_{1} \)}&
{\footnotesize 17.8} &
{\footnotesize 1.1} &
{\footnotesize \( Y^{\pi p}_{1} \)}&
{\footnotesize 60.9 }&
{\footnotesize 2.4 }&
{\footnotesize \( Y^{\pi p}_{1} \)}&
{\footnotesize 29.40} &
{\footnotesize 0.37} \\
{\footnotesize \( Y^{\pi p}_{2} \)}&
{\footnotesize 5.72 }&
{\footnotesize 0.16 }&
{\footnotesize \( Y^{\pi p}_{2} \)}&
{\footnotesize 5.79 }&
{\footnotesize 0.16 }&
{\footnotesize \( Y^{\pi p}_{2} \)}&
{\footnotesize 6.04} &
{\footnotesize 0.26} \\
{\footnotesize \( Y^{Kp}_{1} \)}&
{\footnotesize 5.72 }&
{\footnotesize 1.40} &
{\footnotesize \( Y^{Kp}_{1} \)}&
{\footnotesize 49.3} &
{\footnotesize 2.5 }&
{\footnotesize \( Y^{Kp}_{1} \)}&
{\footnotesize 16.43 }&
{\footnotesize 0.33 }\\
{\footnotesize \( Y^{Kp}_{2} \)}&
{\footnotesize 13.13} &
{\footnotesize 0.38} &
{\footnotesize \( Y^{Kp}_{2} \)}&
{\footnotesize 13.42 }&
{\footnotesize 0.38 }&
{\footnotesize \( Y^{Kp}_{2} \)}&
{\footnotesize 14.07} &
{\footnotesize 0.62 }\\
{\footnotesize \( Y^{\Sigma p}_{1} \)}&
{\footnotesize -250.} &
{\footnotesize 130. }&
{\footnotesize \( Y^{\Sigma p}_{1} \)}&
{\footnotesize 82.4} &
{\footnotesize 6.4 }&
{\footnotesize \( Y^{\Sigma p}_{1} \)}&
{\footnotesize -6. }&
{\footnotesize 35. }\\
{\footnotesize \( Y^{\Sigma p}_{2} \)}&
{\footnotesize -320.} &
{\footnotesize 150.} &
{\footnotesize \( Y^{\Sigma p}_{2} \)}&
{\footnotesize 10. }&
{\footnotesize 22. }&
{\footnotesize \( Y^{\Sigma p}_{2} \)}&
{\footnotesize 72. }&
{\footnotesize 67.} \\
{\footnotesize \( Y^{\gamma p}_{1} \)}&
{\footnotesize 0.0339} &
{\footnotesize 0.0079} &
{\footnotesize \( Y^{\gamma p}_{1} \)}&
{\footnotesize 0.292} &
{\footnotesize 0.013 }&
{\footnotesize \( Y^{\gamma p}_{1} \)}&
{\footnotesize 0.1187} &
{\footnotesize 0.0047 }\\
{\footnotesize \( Y^{\gamma \gamma }_{1} \)}&
{\footnotesize 0.00028 }&
{\footnotesize 0.00015} &
\multicolumn{1}{|c|}{{\footnotesize \( Y^{\gamma \gamma }_{1} \)}}&
\multicolumn{1}{c|}{{\footnotesize 0.000814 }}&
\multicolumn{1}{c||}{{\footnotesize 0.000040} }&
\multicolumn{1}{|c|}{{\footnotesize \( Y^{\gamma \gamma }_{1} \)}}&
\multicolumn{1}{c|}{{\footnotesize 0.00036}}&
\multicolumn{1}{c|}{{\footnotesize 0.00010}}\\
\cline{4-4} \cline{5-5} \cline{6-6} \cline{7-7} \cline{8-8} \cline{9-9}
\multicolumn{1}{|c|}{{\footnotesize \( \delta  \)}}&
\multicolumn{1}{c|}{{\footnotesize 0.00371} }&
\multicolumn{1}{c||}{{\footnotesize 0.00035 }}&
\multicolumn{6}{|c}{}\\
\multicolumn{1}{|c|}{{\footnotesize \( B \)}}&
{\footnotesize 0.3152} &
{\footnotesize 0.0095} &
\multicolumn{6}{|c}{}\\
\cline{1-1} \cline{2-2} \cline{3-3}
\end{tabular}\par}

\medskip{}
{\centering {\small Table 8: parameters of three representative models, defined as in Eq.~(1), for $\sqrt{s}>5$ GeV.}\small
\par}
\bigskip{}

\subsection{Future prospects}

One problem remaining in the analysis of the forward data is the difficulty
in adequately fitting the data for the \( \rho  \) parameter in \( pp \)
and in \( \pi ^{+}p \) reactions. The extraction of the \( \rho  \)
data from the measurements of the differential cross sections data
at small \( t \) is a delicate problem. A re-analysis of these data
may be needed, but it will call for simultaneous fits to the total
cross section data and to the elastic differential cross sections
in the Coulomb-nuclear interference region and in the diffractive
cones, hence an extension of the parametrisation considered here to
the non-forward region. One could also consider a class of analytic
models not incorporated in our fits and ranking procedures, class
in which the rising terms would turn on at some dynamical threshold
\( s_{t} \) (demanding the use of exact dispersion relations), or
add lower trajectories to the existing models. Both approaches would
lead to many extra parameters, and will be the subject of a future
study.

On the other hand, the inclusion of other data may very well allow
one to decide finally amongst the various possibilities. One can go
to deep-inelastic data, but the problem here is that the photon occupies
a special position in Regge theory, and hence the singularities of
DIS amplitudes do not need to be the same as those of hadronic amplitudes.
One can also extend the models to non-forward data and off-diagonal
amplitude such as those of diffractive scattering. Such steps will
involve new parameters associated mainly with form factors, but there
are many data, hence there is the hope that this kind of systematic
study may be generalized, and that in the future we may decide on
the nature of Regge singularities.

Finally, it is our intention to develop the ranking scheme further,
probably along the lines of \cite{Haitun}, and to fine-tune the definition
of indicators, in order that a periodic cross assessments of data
and models be available to the community \cite{COMPETE}.

\section*{Acknowledgements }

The COMPAS group was supported in part by the Russian Foundation for
Basic Research grants RFBR-98-07-90381 and RFBR-01-07-90392. K.K.
is in part supported by the U.S. D.o.E. Contract DE-FG-02-91ER40688-Task
A.

V.V.E. and N.P.T. thank the KIAS president C.W. Kim for the kind invitation
to visit Korea Institute for Advanced Study (KIAS) during the two
weeks of October 2000 where the quality indicators were designed,
discussed and tested (KIAS Report Number P01033); V.V.E. thanks N.N. Nikolaev for the clarifying
the situation with cosmic rays data points; we also thank the Uzhgorod
phenomenology group, and especially Jeno Kontros, for fruitful discussions
of preliminary fit results. K.K. wishes to thank the Korea Institute
of Advanced Science and Yonsei University for the warm hospitality
extended to him during the sabbatical year 2000-01. J.R.C., V.V.E.
and K.K. thank Professor Jean-Eudes Augustin for the hospitality at
LPNHE-University Paris 6, where part of this work was done.

\section*{\noindent {\large Appendix 1. Fits to total cross sections only}}
In this appendix, we present the results for fits to total cross sections
for 33 models, which are variations on the parametrisations referred
to in the main text, following the convention explained after Eq.
\ref{SIG}. Table A1.1 gives our results for the ranking of the models,
according to Eq. \ref{PM}. Table A1.2 gives the values of the quality
indicators associated with each model. Table A1.3 shows the values
of the \( \chi ^{2}/dof \) as a function of energy. The value with
a $-$ exponent indicates that the model has a negative pomeron contribution
in the low-energy region of the fit. The models marked with {*} indicates
that the extrapolation of the $\Sigma p$ cross sections overshoot
the $pp$ or go below $\pi+ p$, or that $C=+1$ residues are negative.
\newpage
{\centering {\small \begin{tabular}{|l|r|r|r|r|r|r|r|r|}
\hline
 {\small Model Code}&
\( P_{A^{M}} \)&
 {\small \( P_{C^{M}_{1}} \)}&
 {\small \( P_{C^{M}_{2}} \)}&
 {\small \( P_{U^{M}} \)}&
 {\small \( P_{R^{M}_{1}} \)}&
 {\small \( P_{R^{M}_{2}} \)}&
 {\small \( P_{S^{M}_{1}} \)}&
 {\small Rank \( P^{M} \)}\\
\hline
 {\small \( {\textrm{RRL}2^{qc}(17)} \)}&
 {\small 54 }&
 {\small 50 }&
 {\small 18 }&
 {\small 56 }&
 {\small 30 }&
 {\small 50 }&
 {\small 40 }&
 {\small 298 }\\
\hline
 {\small \( {(\textrm{RR})^{d}\textrm{ PL}2_{u}}(17) \)}&
 {\small 46 }&
 {\small 58 }&
 {\small 46 }&
 {\small 58 }&
 {\small 30 }&
 {\small 24 }&
 {\small 34 }&
 {\small 296 }\\
\hline
 {\small \( {(\textrm{RR}_{c})^{d}\textrm{ PL}2_{u}}(15) \)}&
 {\small 30 }&
 {\small 42 }&
 {\small 54 }&
 {\small 54 }&
 {\small 46 }&
 {\small 22 }&
 {\small 46 }&
 {\small 294 }\\
\hline
 {\small \( {[\textrm{R}^{qc}\textrm{L}2^{qc}]\textrm{R}_{c} (12) }\)}&
 {\small 14 }&
 {\small 44 }&
 {\small 14 }&
 {\small 50 }&
 {\small 52 }&
 {\small 46 }&
 {\small 58 }&
 {\small 278 }\\
\hline
 {\small \( {\textrm{RRL}2(18)} \)}&
 {\small 52 }&
 {\small 54 }&
 {\small 16 }&
 {\small 44 }&
 {\small 18 }&
 {\small 38 }&
 {\small 44 }&
 {\small 266 }\\
\hline
 {\small \( {(\textrm{RR})^{d}\textrm{ P}^{qc}\textrm{ L}2_{u}}(16) \)}&
 {\small 28 }&
 {\small 52 }&
 {\small 22 }&
 {\small 46 }&
 {\small 38 }&
 {\small 36 }&
 {\small 42 }&
 {\small 264 }\\
\hline
 {\small \( {(\textrm{RR}_{c})^{d}\textrm{ P}^{qc}\textrm{ L}2_{u}}(14) \)}&
 {\small 18 }&
 {\small 26 }&
 {\small 30 }&
 {\small 40 }&
 {\small 55 }&
 {\small 34 }&
 {\small 52 }&
 {\small 255 }\\
\hline
 {\small \( {\textrm{RRE}^{qc}}(17) \)}&
 {\small 50 }&
 {\small 36 }&
 {\small 8 }&
 {\small 48 }&
 {\small 30 }&
 {\small 50 }&
 {\small 30 }&
 {\small 252 }\\
\hline
 {\small \( {\textrm{RR}_{c}\textrm{ L}^{qc}}(15) \)}&
 {\small 24 }&
 {\small 32 }&
 {\small 34 }&
 {\small 32 }&
 {\small 46 }&
 {\small 5 }&
 {\small 54 }&
 {\small 227 }\\
\hline
 {\small \( {\textrm{RR}_{c}\textrm{ E}^{qc}}(15) \)}&
 {\small 22 }&
 {\small 38 }&
 {\small 10 }&
 {\small 52 }&
 {\small 20 }&
 {\small 57 }&
 {\small 22 }&
 {\small 221 }\\
\hline
 {\small \( {\textrm{RR}_{c}\textrm{PL}}(19) \)}&
 {\small 4 }&
 {\small 56 }&
 {\small 56 }&
 {\small 42 }&
 {\small 4 }&
 {\small 0 }&
 {\small 56 }&
 {\small 218 }\\
\hline
 {\small \( {[\textrm{R}^{qc}\textrm{ L}^{qc}]\textrm{ R}}(14) \)}&
 {\small 12 }&
 {\small 48 }&
 {\small 24 }&
 {\small 36 }&
 {\small 55 }&
 {\small 10 }&
 {\small 28 }&
 {\small 213 }\\
\hline
 {\small \( {\textrm{RRL}_{nf}(19)} \)\( ^{*} \)}&
 {\small 57 }&
 {\small 28 }&
 {\small 36 }&
 {\small 10 }&
 {\small 35 }&
 {\small 14 }&
 {\small 32 }&
 {\small 212 }\\
\hline
 {\small \( {\textrm{RRL}^{qc}}(17) \)}&
 {\small 57 }&
 {\small 8 }&
 {\small 32 }&
 {\small 26 }&
 {\small 50 }&
 {\small 16 }&
 {\small 20 }&
 {\small 209 }\\
\hline
 {\small \( {\textrm{RRE}_{nf}(19)} \)}&
 {\small 48 }&
 {\small 40 }&
 {\small 12 }&
 {\small 30 }&
 {\small 10 }&
 {\small 30 }&
 {\small 38 }&
 {\small 208 }\\
\hline
 {\small \( {\textrm{RR}_{c}\textrm{ L}2^{qc}}(15) \)}&
 {\small 32 }&
 {\small 0 }&
 {\small 20 }&
 {\small 4 }&
 {\small 46 }&
 {\small 57 }&
 {\small 48 }&
 {\small 207 }\\
\hline
 {\small \( {\textrm{RRL}2_{nf}(19)} \)}&
 {\small 44 }&
 {\small 34 }&
 {\small 6 }&
 {\small 20 }&
 {\small 10 }&
 {\small 54 }&
 {\small 26 }&
 {\small 194 }\\
\hline
 {\small \( {(\textrm{RR})^{d} P_{nf}L2(20)} \)\( ^{*} \)}&
 {\small 42 }&
 {\small 4 }&
 {\small 58 }&
 {\small 16 }&
 {\small 24 }&
 {\small 32 }&
 {\small 18 }&
 {\small 194 }\\
\hline
 {\small \( {\textrm{RRPE}_{u}}(19) \)}&
 {\small 26 }&
 {\small 46 }&
 {\small 44 }&
 {\small 28 }&
 {\small 10 }&
 {\small 27 }&
 {\small 12 }&
 {\small 193 }\\
\hline
 {\small \( {\textrm{RRPL}2_{u}(19)}^* \)}&
 {\small 36 }&
 {\small 14 }&
 {\small 42 }&
 {\small 14 }&
 {\small 35 }&
 {\small 41 }&
 {\small 8 }&
 {\small 190 }\\
\hline
 {\small \( {(\textrm{RR})^{d}\textrm{ P$_{nf}$L}2_{u}}(19) \)}&
 {\small 40 }&
 {\small 2 }&
 {\small 48 }&
 {\small 22 }&
 {\small 24 }&
 {\small 27 }&
 {\small 24 }&
 {\small 187 }\\
\hline
 {\small \( {\textrm{RRP$_{nf}$L}2_{u}(21)} \)}&
 {\small 38 }&
 {\small 24 }&
 {\small 51 }&
 {\small 6 }&
 {\small 15 }&
 {\small 44 }&
 {\small 4 }&
 {\small 182 }\\
\hline
 {\small \( {[\textrm{R}^{qc}\textrm{ L}^{qc}]\textrm{ R}_{c} (12)}\)}&
 {\small 16 }&
 {\small 16 }&
 {\small 26 }&
 {\small 8 }&
 {\small 58 }&
 {\small 5 }&
 {\small 50 }&
 {\small 179 }\\
\hline
 {\small \( {(\textrm{RR})^{d}\textrm{ }\{\textrm{PL2}\}_{nf}(18)} \)}&
 {\small 20 }&
 {\small 18 }&
 {\small 4 }&
 {\small 38 }&
 {\small 6 }&
 {\small 50 }&
 {\small 36 }&
 {\small 172 }\\
\hline
 {\small \( {\textrm{RRPL}(21)} \)\( ^{*} \)}&
 {\small 8 }&
 {\small 20 }&
 {\small 51 }&
 {\small 34 }&
 {\small 15 }&
 {\small 18 }&
 {\small 16 }&
 {\small 162 }\\
\hline
 {\small \( {\textrm{RRL}(18)} \)}&
 {\small 34 }&
 {\small 10 }&
 {\small 28 }&
 {\small 18 }&
 {\small 41 }&
 {\small 8 }&
 {\small 14 }&
 {\small 153 }\\
\hline
 {\small \( {(\textrm{RR})^{d}\textrm{ PL}(19)} \)}&
 {\small 0 }&
 {\small 12 }&
 {\small 0 }&
 {\small 0 }&
 {\small 0 }&
 {\small 41 }&
 {\small 1 }&
 {\small 54 }\\
\hline
 {\small \( {(\textrm{RR})^{d}\textrm{ P$_{nf}$L}_{u}(18)} \)}&
 {\small 2 }&
 {\small 22 }&
 {\small 2 }&
 {\small 2 }&
 {\small 2 }&
 {\small 20 }&
 {\small 1 }&
 {\small 51}\\
\hline
\end{tabular}}\small \par}

\begin{quote}
Table A1.1: ranking of the 28 models having nonzero area of applicability
amongst the 33 in this paper, following Eq. (\ref{PM}), when only
total cross sections are fitted to.
\end{quote}
{\centering {\small \begin{tabular}{|l|c|c|c|c|c|c|c|}
\hline
\multicolumn{8}{|c|}{ Quality indicators }\\
\hline
{\small Model Code}&
{\small \( A^{M} \)}&
{\small \( C^{M}_{1} \)}&
{\small \( C^{M}_{2} \)}&
{\small \( U^{M} \)}&
{\small \( R^{M}_{1} \)}&
{\small \( R^{M}_{2} \)}&
{\small \( S^{M}_{1} \)}\\
\hline
{\small \( {\textrm{RRL}_{nf}(19)} \)\( ^{*} \)}&
{\small 2.60148}&
{\small 75.54 }&
{\small 94.64 }&
{\small 35.50 }&
{\small 29.05 }&
{\small 0.789 }&
{\small 0.156 }\\
\hline
{\small \( {\textrm{RRL}^{qc}(17)} \)}&
{\small 2.60148}&
{\small 59.26 }&
{\small 94.09 }&
{\small 49.68 }&
{\small 32.28 }&
{\small 0.794 }&
{\small 0.099 }\\
\hline
{\small \( {\textrm{RRL}2^{qc}(17)} \)}&
{\small 2.58120}&
{\small 97.36 }&
{\small 87.91 }&
{\small 131.7 }&
{\small 28.17 }&
{\small 0.941 }&
{\small 0.184 }\\
\hline
{\small \( {\textrm{RRL}2(18)} \)}&
{\small 2.58067}&
{\small 97.52 }&
{\small 87.00 }&
{\small 85.08 }&
{\small 26.68 }&
{\small 0.902 }&
{\small 0.198 }\\
\hline
{\small \( {\textrm{RR E}^{qc}(17)} \)}&
{\small 2.56576}&
{\small 86.15 }&
{\small 79.29 }&
{\small 88.38 }&
{\small 28.17 }&
{\small 0.941 }&
{\small 0.146 }\\
\hline
{\small \( {\textrm{RR E}_{nf}(19)} \)}&
{\small 2.56568}&
{\small 91.45 }&
{\small 80.78 }&
{\small 51.16 }&
{\small 25.35 }&
{\small 0.883 }&
{\small 0.177 }\\
\hline
{\small \( {(\textrm{RR})^{d}\textrm{ PL}2_{u}(17)} \)}&
{\small 2.55303}&
{\small 99.78 }&
{\small 99.67 }&
{\small 184.6 }&
{\small 28.17 }&
{\small 0.875 }&
{\small 0.161 }\\
\hline
{\small \( {\textrm{RRL}2_{nf}(19)} \)}&
{\small 2.54792}&
{\small 81.62 }&
{\small 77.64 }&
{\small 41.85 }&
{\small 25.35 }&
{\small 0.942 }&
{\small 0.143 }\\
\hline
{\small \( {(\textrm{RR})^{d}\textrm{ P$_{nf}$L}2(20)} \)\( ^{*} \)}&
{\small 2.53820}&
{\small 58.94 }&
{\small 99.88 }&
{\small 37.60 }&
{\small 27.67 }&
{\small 0.884 }&
{\small 0.098}\\
\hline
{\small \( {(\textrm{RR})^{d}\textrm{ P$_{nf}$L}2_{u}(19)} \)}&
{\small 2.53154}&
{\small 54.71 }&
{\small 99.72 }&
{\small 44.31 }&
{\small 27.67 }&
{\small 0.877 }&
{\small 0.114 }\\
\hline
{\small \( {\textrm{RRP$_{nf}$L}2_{u}(21)} \)}&
{\small 2.52375}&
{\small 67.76 }&
{\small 99.73 }&
{\small 34.40 }&
{\small 26.41 }&
{\small 0.910 }&
{\small 0.008 }\\
\hline
{\small \( {\textrm{RRPL}2_{u}(19)}^* \)}&
{\small 2.52351}&
{\small 62.59 }&
{\small 99.65 }&
{\small 37.14 }&
{\small 29.05 }&
{\small 0.906 }&
{\small 0.018 }\\
\hline
{\small \( {\textrm{RRL}(18)} \)}&
{\small 2.52103}&
{\small 59.95 }&
{\small 93.52 }&
{\small 39.85 }&
{\small 30.58 }&
{\small 0.693 }&
{\small 0.068 }\\
\hline
{\small \( {\textrm{RR}_{c}\textrm{ L}2^{qc}(15)} \)}&
{\small 2.50642}&
{\small 54.11 }&
{\small 88.31 }&
{\small 26.54 }&
{\small 31.69 }&
{\small 0.952 }&
{\small 0.259 }\\
\hline
{\small \( {(\textrm{RR}_{c})^{d}\textrm{ P L}2_{u}(15)} \)}&
{\small 2.47739}&
{\small 94.20 }&
{\small 99.75 }&
{\small 97.71 }&
{\small 31.69 }&
{\small 0.838 }&
{\small 0.220 }\\
\hline
{\small \( {(\textrm{RR})^{d}\textrm{ P}^{qc}\textrm{ L}2_{u}(16)} \)}&
{\small 2.46789}&
{\small 97.49 }&
{\small 92.53 }&
{\small 87.39 }&
{\small 29.82 }&
{\small 0.900 }&
{\small 0.197 }\\
\hline
{\small \( {\textrm{RRPE}_{u}(19)} \)}&
{\small 2.44915}&
{\small 95.83 }&
{\small 99.66 }&
{\small 49.82 }&
{\small 25.35 }&
{\small 0.877 }&
{\small 0.057 }\\
\hline
{\small \( {\textrm{RR}_{c}\textrm{ L}^{qc}(15)} \)}&
{\small 2.42625}&
{\small 78.91 }&
{\small 94.41 }&
{\small 51.99 }&
{\small 31.69 }&
{\small 0.667 }&
{\small 0.331 }\\
\hline
{\small \( {\textrm{RR}_{c}\textrm{ E}^{qc}(15)} \)}&
{\small 2.39977}&
{\small 89.51 }&
{\small 79.88 }&
{\small 95.81 }&
{\small 27.13 }&
{\small 0.952 }&
{\small 0.104 }\\
\hline
{\small \( {(\textrm{RR})^{d}\{\textrm{PL2}\}_{nf}(18)} \)}&
{\small 2.39430}&
{\small 64.63 }&
{\small 72.14 }&
{\small 70.24 }&
{\small 22.84 }&
{\small 0.941 }&
{\small 0.164 }\\
\hline
{\small \( {(\textrm{RR}_{c})^{d}\textrm{ P}^{qc}\textrm{ L}2_{u}(14)} \)}&
{\small 2.38295}&
{\small 75.32 }&
{\small 93.62 }&
{\small 74.78 }&
{\small 33.80 }&
{\small 0.890 }&
{\small 0.310 }\\
\hline
{\small \( {[\textrm{R}^{qc}\textrm{ L}^{qc}]\textrm{ R}_{c}} \)}&
{\small 2.37016}&
{\small 63.32 }&
{\small 92.89 }&
{\small 34.57 }&
{\small 39.00 }&
{\small 0.667 }&
{\small 0.289 }\\
\hline
{\small \( {[\textrm{R}^{qc}\textrm{ L}2^{qc}]\textrm{ R}_{c} (12)} \)}&
{\small 2.36985}&
{\small 94.28 }&
{\small 83.36 }&
{\small 91.56 }&
{\small 33.38 }&
{\small 0.924 }&
{\small 0.491 }\\
\hline
{\small \( {[\textrm{R}^{qc}\textrm{ L}^{qc}]\textrm{ R}(14)} \)}&
{\small 2.36207}&
{\small 96.86 }&
{\small 92.55 }&
{\small 59.94 }&
{\small 33.80 }&
{\small 0.736 }&
{\small 0.145 }\\
\hline
{\small \( {\textrm{RRPL}(21)} \)\( ^{*} \)}&
{\small 2.18238}&
{\small 64.98 }&
{\small 99.73 }&
{\small 58.88 }&
{\small 26.41 }&
{\small 0.810 }&
{\small 0.082 }\\
\hline
{\small \( {\textrm{RR}_{c}\textrm{PL}(19)} \)}&
{\small 1.93416}&
{\small 99.20 }&
{\small 99.84 }&
{\small 78.20 }&
{\small 21.70 }&
{\small 0.561 }&
{\small 0.372 }\\
\hline
{\small \( {(\textrm{RR})^{d}\textrm{ P$_{nf}$L}_{u}(18)} \)}&
{\small 1.62709}&
{\small 65.46 }&
{\small 65.46 }&
{\small 14.98 }&
{\small 12.11 }&
{\small 0.810 }&
{\small 0.000}\\
\hline
{\small \( {(\textrm{RR})^{d}\textrm{ PL}(19)} \)}&
{\small 1.40760}&
{\small 62.17 }&
{\small 62.17 }&
{\small 14.83 }&
{\small 11.50 }&
{\small 0.906 }&
{\small 0.000}\\
\hline
\end{tabular}}\small \par}

\begin{quote}
Table A1.2: quality indicators of the the 28 models having nonzero
area of applicability amongst the 33 models considered in this paper,
following Eqs. (\ref{A}-\ref{S1}) when only total cross sections
are fitted to.\newpage

{}~\vglue -5cm
\end{quote}
\vspace{-2cm}
{\centering {\small \begin{tabular}{|l|c|c|c|c|c|c|c|c|}
\hline
\multicolumn{9}{|c|}{ {\small \( \chi ^{2}/dof \) vs. \( \sqrt{s_{min}} \) in
GeV }}\\
\hline
{\small \( {\textrm{ModelCode}(\textrm{N}_{par})} \)}&
{\small 3 }&
{\small 4 }&
{\small 5 }&
{\small 6 }&
{\small 7 }&
{\small 8 }&
{\small 9 }&
{\small 10 }\\
\hline
{\small \( {\textrm{RRE}_{nf}(19)} \)}&
{\small 1.38 }&
{\small 1.15 }&
{\small \( \bf 0.91 \)}&
{\small \( \bf 0.87 \)}&
{\small \( \bf 0.89 \)}&
{\small \( \bf 0.90 \)}&
{\small \( \bf 0.93 \)}&
{\small \( \bf 0.91 \)}\\
\hline
{\small \( {\textrm{RRE}^{qc}}(17) \)}&
{\small 1.39 }&
{\small 1.17 }&
{\small \( \bf 0.93 \)}&
{\small \( \bf 0.89 \)}&
{\small \( \bf 0.90 \)}&
{\small \( \bf 0.91 \)}&
{\small \( \bf 0.93 \)}&
{\small \( \bf 0.92 \)}\\
\hline
{\small \( {\textrm{RR}_{c}\textrm{ E}^{qc}}(15) \)}&
{\small 2.37 }&
{\small 1.47 }&
{\small 1.05 }&
{\small \( \bf 0.91 \)}&
{\small \( \bf 0.90 \)}&
{\small \( \bf 0.91 \)}&
{\small \( \bf 0.93 \)}&
{\small \( \bf 0.91 \)}\\
\hline
{\small \( {\textrm{RRL}_{nf}(19)} \)\( ^{*} \)}&
{\small 1.31 }&
{\small \( \bf 0.96^{-} \)}&
{\small \( \bf 0.82 \)}&
{\small \( \bf 0.80 \)}&
{\small \( \bf 0.85 \)}&
{\small \( \bf 0.85 \)}&
{\small \( \bf 0.86 \)}&
{\small \( \bf 0.85 \)}\\
\hline
{\small \( {\textrm{RRL}}(18) \)}&
{\small 1.33 }&
{\small \( \bf 0.98 \)}&
{\small \( \bf 0.85 \)}&
{\small \( \bf 0.83 \)}&
{\small \( \bf 0.87 \)}&
{\small \( \bf 0.87 \)}&
{\small \( \bf 0.87 \)}&
{\small \( \bf 0.86 \)}\\
\hline
{\small \( {\textrm{RRL}^{qc}}(17) \)}&
{\small 1.33 }&
{\small \( \bf 0.99^{-} \)}&
{\small \( \bf 0.85 \)}&
{\small \( \bf 0.83 \)}&
{\small \( \bf 0.87 \)}&
{\small \( \bf 0.87 \)}&
{\small \( \bf 0.87 \)}&
{\small \( \bf 0.85 \)}\\
\hline
{\small \( {\textrm{RR}_{c}\textrm{ L}^{qc}}(15) \)}&
{\small 2.20 }&
{\small 1.22 }&
{\small \( \bf 0.95^{-} \)}&
{\small \( \bf 0.84 \)}&
{\small \( \bf 0.86 \)}&
{\small \( \bf 0.86 \)}&
{\small \( \bf 0.87 \)}&
{\small \( \bf 0.85 \)}\\
\hline
{\small \( {[\textrm{R}^{qc}\textrm{ L}^{qc}]\textrm{ R}}(14) \)}&
{\small 1.44 }&
{\small 1.03 }&
{\small \( \bf 0.88^{-} \)}&
{\small \( \bf 0.85 \)}&
{\small \( \bf 0.89 \)}&
{\small \( \bf 0.87 \)}&
{\small \( \bf 0.88 \)}&
{\small \( \bf 0.87 \)}\\
\hline
{\small \( {[\textrm{R}^{qc}\textrm{ L}^{qc}]\textrm{ R}_{c} (12)}\)}&
{\small 2.20 }&
{\small 1.22 }&
{\small \( \bf 0.95^{-} \)}&
{\small \( \bf 0.84 \)}&
{\small \( 0.86 \)}&
{\small \( 0.86 \)}&
{\small \( \bf 0.87 \)}&
{\small \( \bf 0.85 \)}\\
\hline
{\small \( {\textrm{RRL}2_{nf}(19)} \)}&
{\small 1.45 }&
{\small 1.19 }&
{\small \( \bf 0.94 \)}&
{\small \( \bf 0.90 \)}&
{\small \( \bf 0.91 \)}&
{\small \( \bf 0.91 \)}&
{\small \( \bf 0.94 \)}&
{\small \( \bf 0.92 \)}\\
\hline
{\small \( {\textrm{RRL}2}(18) \)}&
{\small 1.33 }&
{\small 1.05 }&
{\small \( \bf 0.88 \)}&
{\small \( \bf 0.85 \)}&
{\small \( 0.91 \)}&
{\small \( 0.89 \)}&
{\small \( 0.90 \)}&
{\small \( 0.89 \)}\\
\hline
{\small \( {\textrm{RRL}2^{qc}}(17) \)}&
{\small 1.33 }&
{\small 1.06 }&
{\small \( \bf 0.88 \)}&
{\small \( \bf 0.85 \)}&
{\small \( 0.88 \)}&
{\small \( 0.88 \)}&
{\small \( 0.90 \)}&
{\small \( 0.89 \)}\\
\hline
{\small \( {\textrm{RR}_{c}\textrm{ L}2^{qc}}(15) \)}&
{\small 2.28 }&
{\small 1.33 }&
{\small \( \bf 0.99 \)}&
{\small \( \bf 0.87 \)}&
{\small \( \bf 0.87 \)}&
{\small \( \bf 0.88 \)}&
{\small \( \bf 0.90 \)}&
{\small \( \bf 0.89 \)}\\
\hline
{\small \( {[\textrm{R}^{qc}\textrm{ L}2^{qc}]\textrm{ R}_{c}}(12) \)}&
{\small 2.39 }&
{\small 1.38 }&
{\small 1.03 }&
{\small \( \bf 0.89 \)}&
{\small \( \bf 0.90 \)}&
{\small \( \bf 0.89 \)}&
{\small \( \bf 0.91 \)}&
{\small \( \bf 0.91 \)}\\
\hline
{\small \( {(\textrm{RR})^{d}\textrm{ L}^{qc}}(15) \)}&
{\small 2.63 }&
{\small 2.02 }&
{\small 1.37 }&
{\small 1.27 }&
{\small 1.22 }&
{\small 1.21 }&
{\small 1.25 }&
{\small 1.08 }\\
\hline
{\small \( {(\textrm{RR})^{d}\textrm{ PL}}(19) \)}&
{\small 2.34 }&
{\small 1.84 }&
{\small 1.34 }&
{\small 1.24 }&
{\small 1.21 }&
{\small 1.21 }&
{\small 1.22 }&
{\small \( \bf 0.97 \)}\\
\hline
{\small \( {(\textrm{RR})^{d}\textrm{ P}^{qc}\textrm{ E}_{u}}(16) \)}&
{\small 1.44 }&
{\small 1.16 }&
{\small 1.02 }&
{\small 1.01 }&
{\small 1.06 }&
{\small 1.06 }&
{\small 1.05 }&
{\small 1.04 }\\
\hline
{\small \( {(\textrm{RR})^{d}\{\textrm{PL2}\}_{nf}(18)} \)}&
{\small 1.91 }&
{\small 1.56 }&
{\small 1.06 }&
{\small \( \bf 0.97 \)}&
{\small \( \bf 0.95 \)}&
{\small \( \bf 0.95 \)}&
{\small \( \bf 0.99 \)}&
{\small \( \bf 0.94 \)}\\
\hline
{\small \( {\textrm{RRPL}}(21) \)\( ^{*} \)}&
{\small 1.33 }&
{\small \( \bf 0.98^{-} \)}&
{\small \( \bf 0.85^{-} \)}&
{\small \( \bf 0.83^{-} \)}&
{\small \( \bf 0.87 \)}&
{\small \( \bf 0.88 \)}&
{\small \( \bf 0.84 \)}&
{\small \( \bf 0.74 \)}\\
\hline
{\small \( {\textrm{RR}_{c}\textrm{PL}}(19) \)}&
{\small 1.33 }&
{\small \( \bf 0.98^{-} \)}&
{\small \( \bf 0.85^{-} \)}&
{\small \( \bf 0.83^{-} \)}&
{\small \( \bf 0.87^{-} \)}&
{\small \( \bf 0.87^{-} \)}&
{\small \( \bf 0.84 \)}&
{\small \( \bf 0.74 \)}\\
\hline
{\small \( {\textrm{RRPL}_{u,nf}}(20)^{-} \)}&
{\small 2.24 }&
{\small 1.42 }&
{\small 1.14 }&
{\small 1.03 }&
{\small \( \bf 0.97^{-} \)}&
{\small \( \bf 0.91^{-} \)}&
{\small \( \bf 0.84^{-} \)}&
{\small \( \bf 0.74^{-} \)}\\
\hline
{\small \( {\textrm{RRPL}_{u}}(18)^{-} \)}&
{\small 2.24 }&
{\small 1.43 }&
{\small 1.16 }&
{\small 1.05 }&
{\small \( \bf 0.99^{-} \)}&
{\small \( \bf 0.93^{-} \)}&
{\small \( \bf 0.85^{-} \)}&
{\small \( \bf 0.76^{-} \)}\\
\hline
{\small \( {(\textrm{RR})^{d}\textrm{ P$_{nf}$L}_{u}}(18) \)}&
{\small 2.66 }&
{\small 2.10 }&
{\small 1.73 }&
{\small 1.58 }&
{\small 1.43 }&
{\small 1.37 }&
{\small 1.25 }&
{\small \( \bf 0.96 \)}\\
\hline
{\small \( {(\textrm{RR})^{d}\textrm{ P}^{qc}\textrm{ L}_{u}}(15) \)}&
{\small 2.74 }&
{\small 2.27 }&
{\small 2.06 }&
{\small 2.06 }&
{\small 2.12 }&
{\small 2.15 }&
{\small 2.19 }&
{\small 2.38 }\\
\hline
{\small \( {(\textrm{RR})^{d}\textrm{ P$_{nf}$ L}2}(20) \)\( ^{*} \)}&
{\small 1.24 }&
{\small \( \bf 0.99 \)}&
{\small \( \bf 0.82 \)}&
{\small \( \bf 0.79 \)}&
{\small \( \bf 0.83 \)}&
{\small \( \bf 0.84 \)}&
{\small \( \bf 0.83 \)}&
{\small \( \bf 0.73 \)}\\
\hline
{\small \( {\textrm{RRPL}2_{u}}(21) \)}&
{\small 1.26 }&
{\small \( \bf 0.97 \)}&
{\small \( \bf 0.81 \)}&
{\small \( \bf 0.79 \)}&
{\small \( \bf 0.82 \)}&
{\small \( \bf 0.83 \)}&
{\small \( \bf 0.82 \)}&
{\small \( \bf 0.75 \)}\\
\hline
{\small \( {\textrm{RRPL}2_{u}}(19)^* \)}&
{\small 1.27 }&
{\small \( \bf 0.98 \)}&
{\small \( \bf 0.82 \)}&
{\small \( \bf 0.80 \)}&
{\small \( \bf 0.84 \)}&
{\small \( \bf 0.84 \)}&
{\small \( \bf 0.83 \)}&
{\small \( \bf 0.76 \)}\\
\hline
{\small \( {(\textrm{RR})^{d}\textrm{ P$_{nf}$ L}2_{u}}(19) \)}&
{\small 1.27 }&
{\small \( \bf 0.99 \)}&
{\small \( \bf 0.82 \)}&
{\small \( \bf 0.80 \)}&
{\small \( \bf 0.83 \)}&
{\small \( \bf 0.83 \)}&
{\small \( \bf 0.82 \)}&
{\small \( \bf 0.75 \)}\\
\hline
{\small \( {(\textrm{RR})^{d}\textrm{ P L}2_{u}}(17) \)}&
{\small 1.28 }&
{\small 1.00 }&
{\small \( \bf 0.82 \)}&
{\small \( \bf 0.81 \)}&
{\small \( \bf 0.83 \)}&
{\small \( \bf 0.83 \)}&
{\small \( \bf 0.83 \)}&
{\small \( \bf 0.76 \)}\\
\hline
{\small \( {(\textrm{RR})^{d}\textrm{ P}^{qc}\textrm{ L}2_{u}}(16) \)}&
{\small 1.30 }&
{\small 1.04 }&
{\small \( \bf 0.88 \)}&
{\small \( \bf 0.87 \)}&
{\small \( \bf 0.91 \)}&
{\small \( \bf 0.91 \)}&
{\small \( \bf 0.90 \)}&
{\small \( \bf 0.86 \)}\\
\hline
{\small \( {(\textrm{RR}_{c})^{d}\textrm{ P L}2_{u}}(15) \)}&
{\small 2.08 }&
{\small 1.19 }&
{\small \( \bf 0.90 \)}&
{\small \( \bf 0.82 \)}&
{\small \( \bf 0.83 \)}&
{\small \( \bf 0.83 \)}&
{\small \( \bf 0.82 \)}&
{\small \( \bf 0.75 \)}\\
\hline
{\small \( {(\textrm{RR}_{c})^{d}\textrm{ P}^{qc}\textrm{ L}2_{u}}(14) \)}&
{\small 2.11 }&
{\small 1.22 }&
{\small \( \bf 0.96 \)}&
{\small \( \bf 0.88 \)}&
{\small \( \bf 0.90 \)}&
{\small \( \bf 0.90 \)}&
{\small \( \bf 0.89 \)}&
{\small \( \bf 0.86 \)}\\
\hline
{\small \( {\textrm{RRPE}_{u}}(19) \)}&
{\small 1.36 }&
{\small 1.04 }&
{\small \( \bf 0.89 \)}&
{\small \( \bf 0.86 \)}&
{\small \( \bf 0.87 \)}&
{\small \( \bf 0.86 \)}&
{\small \( \bf 0.83 \)}&
{\small \( \bf 0.76 \) }\\
\hline
\end{tabular}}\small \par}

\begin{quote}
Table A1.3: \( \chi ^{2}/dof \) as a function of the minimum energy
of the fit for the 33 models considered in this paper when only total
cross sections are fitted to.
\end{quote}
{\centering \newpage\par}

\noindent \textbf{\large Appendix 2. Fits to total cross sections
and to the \( \rho  \) parameter. }{\large \par}

\noindent In this appendix, we present the results for fits to total
cross sections and the \( \rho  \) parameter for 33 models, which
are variations on the parametrisations referred to in the main text,
following the convention explained after Eq. \ref{SIG}. Only 21 of
these passed through qualification tests in this case. The tables
are presented as in Appendix 1. It should be noted that for model
RRPL2\( _{u} \)(19) with highest rank, corresponding to model RRP$_{nf}$L2\( _{u}
\)(21)
with the extra imposition of factorization on the \( P_{ab}  \)
residues, tends to choose a negative value for the reggeon \( C=+1 \)
residue in \( \gamma \gamma  \) cross sections. Although this does
not exclude it as the residue has large errors, we have preferred
to present in this paper the details of the next best ranking
parametrisation.\newpage

{\centering \begin{tabular}{|l|r|r|r|r|r|r|r|r|r|}
\hline
 Model Code&
\( P_{A^{M}} \)&
 \( P_{C^{M}_{1}} \)&
 \( P_{C^{M}_{2}} \)&
 \( P_{U^{M}} \)&
 \( P_{R^{M}_{1}} \)&
 \( P_{R^{M}_{2}} \)&
 \( P_{S^{M}_{1}} \)&
\( P_{S^{M}_{2}} \)&
 Rank \( P^{M} \)\\
\hline
 \( {\textrm{RRPL}2_{u}(19)}^* \)&
 42 &
 26 &
 42 &
 42 &
 34 &
 28 &
 12 &
 4 &
 230 \\
\hline
 \( {\textrm{RRP$_{nf}$L}2_{u}(21)} \)&
 44 &
 36 &
 44 &
 40 &
 15 &
 31 &
 10 &
 2 &
 222 \\
\hline
\( {\textrm{RRL}_{nf}(19)} \){\small \( ^{*} \)}&
 30 &
 42 &
 26 &
 24 &
 34 &
 18 &
 18 &
 30 &
 222 \\
\hline
 \( {(\textrm{RR}_{c})^{d}\textrm{ PL}2_{u}}(15) \)&
 34 &
 20 &
 36 &
 20 &
 28 &
 24 &
 28 &
 14 &
 204 \\
\hline
 \( {(\textrm{RR})^{d}\textrm{ P$_{nf}$L}2_{u}}(19) \)&
 40 &
 8 &
 40 &
 22 &
 34 &
 22 &
 16 &
 12 &
 194 \\
\hline
 \( {[\textrm{R}^{qc}\textrm{ L}^{qc}]\textrm{R}_{c}}(12) \)&
 14 &
 32 &
 18 &
 10 &
 42 &
 6 &
 24 &
 38 &
 184 \\
\hline
 \( {(\textrm{RR}_{c})^{d}\textrm{P}^{qc}\textrm{L}2_{u}}(14) \)&
 20 &
 16 &
 10 &
 36 &
 19 &
 36 &
 22 &
 22 &
 181 \\
\hline
 \( {(\textrm{RR})^{d}\textrm{P}^{qc}\textrm{L}2_{u}}(16) \)&
 18 &
 14 &
 8 &
 38 &
 8 &
 38 &
 30 &
 26 &
 180 \\
\hline
 \( {\textrm{RR}_{c}\textrm{ L}2^{qc}(15)} \)&
 6 &
 30 &
 6 &
 4 &
 6 &
 44 &
 44 &
 40 &
 180 \\
\hline
 \( {(\textrm{RR})^{d}\textrm{ P$_{nf}$L}2(20)} \){\small \( ^{*} \)}&
 38 &
 2 &
 28 &
 32 &
 25 &
 31 &
 14 &
 8 &
 178 \\
\hline
 \( {(\textrm{RR})^{d}\textrm{ PL}2_{u}}(17) \)&
 36 &
 0 &
 34 &
 18 &
 30 &
 26 &
 20 &
 10 &
 174 \\
\hline
 \( {\textrm{RRPL}(21)} \){\small \( ^{*} \)}&
 2 &
 34 &
 32 &
 44 &
 15 &
 16 &
 6 &
 24 &
 173 \\
\hline
 \( {\textrm{RR}_{c}\textrm{ L}^{qc}}(15) \)&
 24 &
 38 &
 24 &
 8 &
 10 &
 4 &
 32 &
 32 &
 172 \\
\hline
 \( {\textrm{RRL}2^{qc}(17)} \)&
 10 &
 28 &
 4 &
 2 &
 2 &
 42 &
 40 &
 42 &
 170 \\
\hline
 \( {[\textrm{R}^{qc}\textrm{ L}2^{qc}]\textrm{R}_{c}}(12) \)&
 12 &
 18 &
 0 &
 6 &
 22 &
 40 &
 38 &
 34 &
 170 \\
\hline
 \( {\textrm{RRL}^{qc}(17)} \)&
 28 &
 6 &
 20 &
 30 &
 44 &
 12 &
 4 &
 18 &
 162 \\
\hline
 \( {\textrm{RRPE}_{u}}(19) \)&
 22 &
 44 &
 12 &
 16 &
 4 &
 20 &
 34 &
 6 &
 158 \\
\hline
 \( {[\textrm{R}^{qc}\textrm{L}^{qc}]\textrm{R}}(14) \)&
 16 &
 24 &
 14 &
 12 &
 19 &
 14 &
 36 &
 20 &
 155 \\
\hline
 \( {\textrm{RRL}2(18)} \)&
 8 &
 22 &
 2 &
 0 &
 0 &
 34 &
 42 &
 44 &
 152 \\
\hline
 \( {\textrm{RR}_{c}\textrm{PL}}(19) \)&
 4 &
 12 &
 38 &
 14 &
 12 &
 0 &
 26 &
 36 &
 142 \\
\hline
 \( {\textrm{RRL}(18)} \)&
 26 &
 10 &
 16 &
 26 &
 39 &
 8 &
 8 &
 0 &
133\\
\hline
\end{tabular}\par}

\begin{quote}
Table A2.1: ranking of the the 21 models having nonzero area of applicability
amongst the 33 models considered in this paper, following Eq. (\ref{PM})
when cross sections and \( \rho  \) parameters are fitted to.

\end{quote}
{\noindent \centering {\small \begin{tabular}{|l|c|c|c|c|c|c|c|c|}
\hline
\multicolumn{9}{|l|}{ Quality indicators }\\
\hline
{\small Model Code}&
{\small \( A^{M} \)}&
{\small \( C^{M}_{1} \)}&
{\small \( C^{M}_{2} \)}&
{\small \( U^{M} \)}&
{\small \( R^{M}_{1} \)}&
{\small \( R^{M}_{2} \)}&
{\small \( S^{M}_{1} \)}&
{\small \( S^{M}_{2} \)}\\
\hline
{\small \( {\textrm{RRP$_{nf}$L}2_{u}(21)} \)}&
{\small 2.20661}&
{\small 67.98 }&
{\small 84.74 }&
{\small 22.88 }&
{\small 29.45 }&
{\small 0.900 }&
{\small 0.224 }&
{\small 0.101 }\\
\hline
{\small \( {\textrm{RRPL}2_{u}(19)}^* \)}&
{\small 2.20619}&
{\small 63.46 }&
{\small 84.13 }&
{\small 24.14 }&
{\small 32.40 }&
{\small 0.895 }&
{\small 0.226 }&
{\small 0.190 }\\
\hline
{\small \( {(\textrm{RR})^{d}\textrm{ P$_{nf}$L}2_{u}(19)} \)}&
{\small 2.18781}&
{\small 53.15 }&
{\small 83.81 }&
{\small 16.49 }&
{\small 32.40 }&
{\small 0.871 }&
{\small 0.286 }&
{\small 0.690 }\\
\hline
{\small \( {(\textrm{RR})^{d}\textrm{ P$_{nf}$L}2(20)} \)\( ^{*} \)}&
{\small 2.18530}&
{\small 50.41 }&
{\small 81.74 }&
{\small 18.21 }&
{\small 30.86 }&
{\small 0.900 }&
{\small 0.265 }&
{\small 0.407 }\\
\hline
{\small \( {(\textrm{RR})^{d}\textrm{ PL}2_{u}}(17) \)}&
{\small 1.99653}&
{\small 50.35 }&
{\small 83.04 }&
{\small 15.64 }&
{\small 31.61 }&
{\small 0.882 }&
{\small 0.296 }&
{\small 0.673 }\\
\hline
{\small \( {(\textrm{RR}_{c})^{d}\textrm{ PL}2_{u}}(15) \)}&
{\small 1.88491}&
{\small 61.92 }&
{\small 83.38 }&
{\small 16.26 }&
{\small 31.13 }&
{\small 0.876 }&
{\small 0.467 }&
{\small 0.795 }\\
\hline
{\small \( {\textrm{RRL}_{nf}(19)} \)\( ^{*} \)}&
{\small 1.82464}&
{\small 73.37 }&
{\small 81.09 }&
{\small 16.63 }&
{\small 32.40 }&
{\small 0.784 }&
{\small 0.289 }&
{\small 1.302 }\\
\hline
{\small \( {\textrm{RRL}^{qc}(17)} \)}&
{\small 1.82281}&
{\small 52.97 }&
{\small 78.17 }&
{\small 17.56 }&
{\small 36.00 }&
{\small 0.743 }&
{\small 0.198 }&
{\small 1.080 }\\
\hline
{\small \( {\textrm{RRL}(18)} \)}&
{\small 1.82274}&
{\small 53.59 }&
{\small 77.18 }&
{\small 16.73 }&
{\small 34.11 }&
{\small 0.686 }&
{\small 0.217 }&
{\small 0.001 }\\
\hline
{\small \( {\textrm{RR}_{c}\textrm{ L}^{qc}}(15) \)}&
{\small 1.82270}&
{\small 68.31 }&
{\small 79.68 }&
{\small 12.48 }&
{\small 28.31 }&
{\small 0.667 }&
{\small 0.525 }&
{\small 1.311 }\\
\hline
{\small \( {\textrm{RRPE}_{u}}(19) \)}&
{\small 1.81878}&
{\small 73.98 }&
{\small 73.74 }&
{\small 15.46 }&
{\small 22.65 }&
{\small 0.830 }&
{\small 0.526 }&
{\small 0.282 }\\
\hline
{\small \( {(\textrm{RR}_{c})^{d}\textrm{P}^{qc}\textrm{L}2_{u}}(14) \)}&
{\small 1.79558}&
{\small 60.29 }&
{\small 67.08 }&
{\small 19.94 }&
{\small 30.20 }&
{\small 0.912 }&
{\small 0.429 }&
{\small 1.100 }\\
\hline
{\small \( {(\textrm{RR})^{d}\textrm{P}^{qc}\textrm{L}2_{u}}(16) \)}&
{\small 1.79315 }&
{\small 58.40 }&
{\small 66.41 }&
{\small 19.98 }&
{\small 26.65 }&
{\small 0.917 }&
{\small 0.470 }&
{\small 1.241 }\\
\hline
{\small \( {[\textrm{R}^{qc}\textrm{L}^{qc}]\textrm{R}}(14) \)}&
{\small 1.73409}&
{\small 63.29 }&
{\small 76.41 }&
{\small 13.09 }&
{\small 30.20 }&
{\small 0.747 }&
{\small 0.533 }&
{\small 1.082 }\\
\hline
{\small \( {[\textrm{R}^{qc}\textrm{ L}^{qc}]\textrm{R}_{c}}(12) \)}&
{\small 1.73264}&
{\small 65.79 }&
{\small 78.13 }&
{\small 13.03 }&
{\small 34.85 }&
{\small 0.682 }&
{\small 0.440 }&
{\small 1.935 }\\
\hline
{\small \( {[\textrm{R}^{qc}\textrm{ L}2^{qc}]\textrm{ R}_{c} (12)} \)}&
{\small 1.72644}&
{\small 61.50 }&
{\small 61.50 }&
{\small 11.58 }&
{\small 30.54 }&
{\small 0.939 }&
{\small 1.159 }&
{\small 1.692 }\\
\hline
{\small \( {\textrm{RRL}2^{qc}(17)} \)}&
{\small 1.72618}&
{\small 64.20 }&
{\small 64.20 }&
{\small 11.23 }&
{\small 22.06 }&
{\small 0.941 }&
{\small 1.318 }&
{\small 2.503 }\\
\hline
{\small \( {\textrm{RRL}2(18)} \)}&
{\small 1.72607}&
{\small 63.04 }&
{\small 63.04 }&
{\small 11.19 }&
{\small 20.89 }&
{\small 0.902 }&
{\small 1.395 }&
{\small 2.657 }\\
\hline
{\small \( {\textrm{RR}_{c}\textrm{ L}2^{qc}(15)} \)}&
{\small 1.72369}&
{\small 65.63 }&
{\small 65.63 }&
{\small 11.27 }&
{\small 24.81 }&
{\small 0.952 }&
{\small 1.447 }&
{\small 2.104 }\\
\hline
{\small \( {\textrm{RR}_{c}\textrm{PL}}(19) \)}&
{\small 1.99062}&
{\small 55.13 }&
{\small 83.67 }&
{\small 15.38 }&
{\small 28.45 }&
{\small 0.61 }&
{\small 0.466 }&
{\small 1.824 }\\
\hline
{\small \( {\textrm{RRPL}(21)} \)\( ^{*} \)}&
{\small 1.60724}&
{\small 66.59 }&
{\small 82.16 }&
{\small 26.29 }&
{\small 29.45 }&
{\small 0.752 }&
{\small 0.210 }&
{\small 1.135 }\\
\hline
\end{tabular}}\small \par}

\begin{quote}
Table A2.2: quality indicators of the 21 models having nonzero area
of applicability amongst the 33 models considered in this paper, following
Eqs. (\ref{A}-\ref{S1}) and (\ref{S2}) when cross sections and
\( \rho  \) parameters are fitted to.
\end{quote}
\newpage

{}~\vglue -2cm

{\centering {\small \begin{tabular}{|l|c|c|c|c|c|c|c|c|}
\hline
\multicolumn{9}{|c|}{ {\small \( \chi ^{2}/dof \) vs. \( \sqrt{s_{min}} \) in
GeV }}\\
\hline
{\small \( {\textrm{Model Code}(\textrm{N}_{par})} \)}&
{\small 3 }&
{\small 4 }&
{\small 5 }&
{\small 6 }&
{\small 7 }&
{\small 8 }&
{\small 9 }&
{\small 10 }\\
\hline
{\small \( {\textrm{RRE}_{nf}(19)} \)}&
{\small 1.83 }&
{\small 1.38 }&
{\small 1.12 }&
{\small 1.10 }&
{\small 1.10 }&
{\small 1.05 }&
{\small 1.02 }&
{\small 1.02 }\\
\hline
{\small \( {\textrm{RRE}^{qc}}(17) \)}&
{\small 1.84 }&
{\small 1.39 }&
{\small 1.13 }&
{\small 1.12 }&
{\small 1.11 }&
{\small 1.06 }&
{\small 1.02 }&
{\small 1.02 }\\
\hline
{\small \( {\textrm{RR}_{c}\textrm{ E}^{qc}}(15) \)}&
{\small 2.47 }&
{\small 1.58 }&
{\small 1.23 }&
{\small 1.13 }&
{\small 1.10 }&
{\small 1.05 }&
{\small 1.02 }&
{\small 1.02 }\\
\hline
{\small \( {\textrm{RRL}_{nf}(19)} \)\( ^{*} \)}&
{\small 1.61 }&
{\small 1.10 }&
{\small \( \bf 0.97^{-} \)}&
{\small \( \bf 0.97^{-} \)}&
{\small 1.00 }&
{\small \( \bf 0.96 \)}&
{\small \( \bf 0.94 \)}&
{\small \( \bf 0.93 \)}\\
\hline
{\small \( {\textrm{RRL}}(18) \)}&
{\small 1.63 }&
{\small 1.13 }&
{\small \( \bf 0.99^{-} \)}&
{\small \( \bf 0.99^{-} \)}&
{\small 1.02 }&
{\small \( \bf 0.97 \)}&
{\small \( \bf 0.95 \)}&
{\small \( \bf 0.94 \)}\\
\hline
{\small \( {\textrm{RRL}^{qc}}(17) \)}&
{\small 1.63 }&
{\small 1.13 }&
{\small \( \bf 1.00^{-} \)}&
{\small \( \bf 0.99^{-} \)}&
{\small 1.02 }&
{\small \( \bf 0.97 \)}&
{\small \( \bf 0.94 \)}&
{\small \( \bf 0.94 \)}\\
\hline
{\small \( {\textrm{RR}_{c}\textrm{ L}^{qc}}(15) \)}&
{\small 2.20 }&
{\small 1.30 }&
{\small 1.08 }&
{\small 1.01 }&
{\small 1.02 }&
{\small \( \bf 0.97 \)}&
{\small \( \bf 0.94 \)}&
{\small \( \bf 0.94 \)}\\
\hline
{\small \( {[\textrm{R}^{qc}\textrm{ L}^{qc}]\textrm{ R}}(14) \)}&
{\small 1.70 }&
{\small 1.16 }&
{\small 1.02 }&
{\small 1.01 }&
{\small 1.03 }&
{\small \( \bf 0.98^{-} \)}&
{\small \( \bf 0.95 \)}&
{\small \( \bf 0.94 \)}\\
\hline
{\small \( {[\textrm{R}^{qc}\textrm{ L}^{qc}]\textrm{ R}_{c}}(12) \)}&
{\small 2.20 }&
{\small 1.30 }&
{\small 1.08 }&
{\small 1.01 }&
{\small 1.02 }&
{\small \( \bf 0.97^{-} \)}&
{\small \( \bf 0.94 \)}&
{\small \( \bf 0.94 \)}\\
\hline
{\small \( {\textrm{RRL}2_{nf}(19)} \)}&
{\small 1.83 }&
{\small 1.34 }&
{\small 1.11 }&
{\small 1.10 }&
{\small 1.11 }&
{\small 1.06 }&
{\small 1.01 }&
{\small 1.00 }\\
\hline
{\small \( {\textrm{RRL}2}(18) \)}&
{\small 1.68 }&
{\small 1.22 }&
{\small 1.04 }&
{\small 1.04 }&
{\small 1.06 }&
{\small 1.01 }&
{\small \( \bf 0.97 \)}&
{\small \( \bf 0.97 \)}\\
\hline
{\small \( {\textrm{RRL}2^{qc}}(17) \)}&
{\small 1.68 }&
{\small 1.22 }&
{\small 1.04 }&
{\small 1.04 }&
{\small 1.05 }&
{\small 1.01 }&
{\small \( \bf 0.97 \)}&
{\small \( \bf 0.97 \)}\\
\hline
{\small \( {\textrm{RR}_{c}\textrm{ L}2^{qc}}(15) \)}&
{\small 2.30 }&
{\small 1.41 }&
{\small 1.13 }&
{\small 1.06 }&
{\small 1.05 }&
{\small 1.00 }&
{\small \( \bf 0.97 \)}&
{\small \( \bf 0.97 \)}\\
\hline
{\small \( {[\textrm{R}^{qc}\textrm{ L}2^{qc}]\textrm{ R}_{c}}(12) \)}&
{\small 2.38 }&
{\small 1.44 }&
{\small 1.16 }&
{\small 1.07 }&
{\small 1.07 }&
{\small 1.01 }&
{\small \( \bf 0.98 \)}&
{\small \( \bf 0.98 \)}\\
\hline
{\small \( {(\textrm{RR})^{d}\textrm{ L}^{qc}}(15) \)}&
{\small 3.76 }&
{\small 2.61 }&
{\small 1.87 }&
{\small 1.82 }&
{\small 1.73 }&
{\small 1.70 }&
{\small 1.72 }&
{\small 1.72 }\\
\hline
{\small \( {(\textrm{RR})^{d}\textrm{ PL}}(19) \)}&
{\small 3.45 }&
{\small 2.37 }&
{\small 1.81 }&
{\small 1.76 }&
{\small 1.71 }&
{\small 1.69 }&
{\small 1.73 }&
{\small 1.72 }\\
\hline
{\small \( {(\textrm{RR})^{d}\textrm{ P}^{qc}\textrm{ E}_{u}}(16) \)}&
{\small 2.35 }&
{\small 1.53 }&
{\small 1.24 }&
{\small 1.23 }&
{\small 1.21 }&
{\small 1.17 }&
{\small 1.17 }&
{\small 1.17 }\\
\hline
{\small \( {(\textrm{RR})^{d}\{\textrm{PL2}\}_{nf}}(18) \)}&
{\small 2.81 }&
{\small 1.98 }&
{\small 1.40 }&
{\small 1.34 }&
{\small 1.27 }&
{\small 1.20 }&
{\small 1.13 }&
{\small 1.12 }\\
\hline
{\small \( {\textrm{RRPL}}(21) \)\( ^{*} \)}&
{\small 1.63 }&
{\small 1.11 }&
{\small \( \bf 0.98^{-} \)}&
{\small \( \bf 0.98^{-} \)}&
{\small \( \bf 0.99^{-} \)}&
{\small \( \bf 0.94^{-} \)}&
{\small \( \bf 0.93^{-} \)}&
{\small \( \bf 0.91 \)}\\
\hline
{\small \( {\textrm{RR}_{c}\textrm{PL}}(19) \)}&
{\small 1.63 }&
{\small 1.11 }&
{\small \( \bf 0.98^{-} \)}&
{\small \( \bf 0.98^{-} \)}&
{\small \( \bf 0.99^{-} \)}&
{\small \( \bf 0.94^{-} \)}&
{\small \( \bf 0.93^{-} \)}&
{\small \( \bf 0.91 \)}\\
\hline
{\small \( {\textrm{RRPL}_{u,nf}}(20)^{-} \)}&
{\small 2.43 }&
{\small 1.49 }&
{\small 1.25 }&
{\small 1.16 }&
{\small 1.08 }&
{\small \( 1.00^{-} \)}&
{\small \( \bf 0.97^{-} \)}&
{\small \( \bf 0.92^{-} \)}\\
\hline
{\small \( {\textrm{RRPL}_{u}}(18)^{-} \)}&
{\small 2.43 }&
{\small 1.50 }&
{\small 1.27 }&
{\small 1.17 }&
{\small 1.10 }&
{\small 1.01 }&
{\small \( \bf 0.98^{-} \)}&
{\small \( \bf 0.93^{-} \)}\\
\hline
{\small \( {(\textrm{RR})^{d}\textrm{ P$_{nf}$L}_{u}}(18) \)}&
{\small 3.59 }&
{\small 2.50 }&
{\small 2.10 }&
{\small 1.95 }&
{\small 1.91 }&
{\small 1.88 }&
{\small 1.89 }&
{\small 1.87 }\\
\hline
{\small \( {(\textrm{RR})^{d}\textrm{ P}^{qc}\textrm{ L}_{u}}(15) \)}&
{\small 3.67 }&
{\small 2.64 }&
{\small 2.32 }&
{\small 2.27 }&
{\small 2.32 }&
{\small 2.32 }&
{\small 2.39 }&
{\small 2.51 }\\
\hline
{\small \( {(\textrm{RR})^{d}\textrm{ P$_{nf}$ L}2}(20) \)\( ^{*} \)}&
{\small 1.92 }&
{\small 1.23 }&
{\small \( \bf 1.00 \)}&
{\small \( \bf 1.00 \)}&
{\small \( \bf 0.99 \)}&
{\small \( \bf 0.94 \)}&
{\small \( \bf 0.93 \)}&
{\small \( \bf 0.92 \)}\\
\hline
{\small \( {\textrm{RRP$_{nf}$L}2_{u}}(21) \)}&
{\small 1.75 }&
{\small 1.14 }&
{\small \( \bf 0.97 \)}&
{\small \( \bf 0.97 \)}&
{\small \( \bf 0.97 \)}&
{\small \( \bf 0.92 \)}&
{\small \( \bf 0.93 \)}&
{\small \( \bf 0.92 \)}\\
\hline
{\small \( {\textrm{RRPL}2_{u}}(19)^* \)}&
{\small 1.75 }&
{\small 1.15 }&
{\small \( \bf 0.98 \)}&
{\small \( \bf 0.98 \)}&
{\small \( \bf 0.97 \)}&
{\small \( \bf 0.93 \)}&
{\small \( \bf 0.93 \)}&
{\small \( \bf 0.92 \)}\\
\hline
{\small \( {(\textrm{RR})^{d}\textrm{ P$_{nf}$ L}2_{u}}(19) \)}&
{\small 1.96 }&
{\small 1.26 }&
{\small \( \bf 0.99 \)}&
{\small \( \bf 0.99 \)}&
{\small \( \bf 0.98 \)}&
\textbf{\small \( \bf 0.93 \)}&
{\small \( \bf 0.93 \)}&
{\small \( \bf 0.93 \)}\\
\hline
{\small \( {(\textrm{RR})^{d}\textrm{ PL}2_{u}}(17) \)}&
{\small 1.96 }&
{\small 1.27 }&
{\small 1.00 }&
{\small \( \bf 1.00 \)}&
{\small \( \bf 0.98 \)}&
{\small \( \bf 0.94 \)}&
{\small \( \bf 0.93 \)}&
{\small \( \bf 0.93 \)}\\
\hline
{\small \( {(\textrm{RR})^{d}\textrm{ P}^{qc}\textrm{ L}2_{u}}(16) \)}&
{\small 1.98 }&
{\small 1.29 }&
{\small 1.04 }&
{\small 1.04 }&
{\small 1.03 }&
{\small \( \bf 0.98 \)}&
{\small \( \bf 0.97 \)}&
{\small \( \bf 0.97 \)}\\
\hline
{\small \( {(\textrm{RR}_{c})^{d}\textrm{ PL}2_{u}}(15) \)}&
{\small 2.38 }&
{\small 1.37 }&
{\small 1.06 }&
{\small 1.01 }&
{\small \( \bf 0.98 \)}&
{\small \( \bf 0.93 \)}&
{\small \( \bf 0.93 \)}&
{\small \( \bf 0.93 \)}\\
\hline
{\small \( {(\textrm{RR}_{c})^{d}\textrm{ P}^{qc}\textrm{ L}2_{u}}(14) \)}&
{\small 2.40 }&
{\small 1.39 }&
{\small 1.10 }&
{\small 1.05 }&
{\small 1.03 }&
{\small \( \bf 0.98 \)}&
{\small \( \bf 0.97 \)}&
{\small \( \bf 0.97 \)}\\
\hline
{\small \( {\textrm{RRPE}_{u}}(19) \)}&
{\small 1.88 }&
{\small 1.22 }&
{\small 1.06 }&
{\small 1.03 }&
{\small 1.01 }&
{\small \( \bf 0.96 \)}&
\textbf{\small \( \bf 0.95 \)}&
{\small \( \bf 0.93 \) }\\
\hline
\end{tabular}}\small \par}

\begin{quote}
Table A2.3: \( \chi ^{2}/dof \) as a function of the minimum energy
of the fit for the 33 models considered in this paper when cross sections
and \( \rho  \) parameters are fitted to.
\end{quote}
\section*{\noindent {\large Appendix 3. Formul\ae}}
We give here the formulae used in this paper. The imaginary part
of the amplitude, which we take as $s$ time the total cross section,
is parametrized as the sum of several terms, $I_n$, with 
(see Eq.~(\ref{SIG})):
\begin{eqnarray}
I^+_{pole}      &=&C^+ (s/s_{1})^{\alpha_{+}}\\
I^-_{pole}      &=&\mp C^- (s/s_{1})^{\alpha_{-}}\\
I_{L}   &=& C_L\ s \ln (s/s_{1})\\
I_{L2}  &=&C_{L2}\ s \ln ^{2}(s/s_{0})
\end{eqnarray}
All terms have charge conjugation $C=+1$, except $I^-_{pole}$ which has $C=-1$.
We can obtain the corresponding additive 
real parts through $s$ to $u$ crossing symmetry and analyticity:
\begin{eqnarray}
R^+_{pole}   &=&-I^+_{pole}\cot\left[{\pi\over 2}\ \alpha_+\right]\\
R^-_{pole}   &=& I^-_{pole} \tan\left[{\pi\over 2}\ \alpha_-\right]\\
R_{L}&=&{\pi\over 2}\ s\ C_L\\
R_{L2}&=&\pi\  s\ \ln\left({s/s_0}\right)\ C_{L2}
\end{eqnarray}

\begin{thebibliography}{10}
\bibitem{Froissart}M. Froissart,  \emph{Phys. Rev.} {\bf 123}, 1053 (1961).
%%CITATION = PHRVA,123,1053;%%,
A. Martin, \emph{Nuovo Cimento} \textbf{42}, 930 (1966).
\bibitem{Heisenberg:1952}W.~Heisenberg, \emph{Zeit. Phys.} \textbf{133}, 
65 (1952). (In German)
\bibitem{Martin}L. Lukaszuk and A. Martin, \emph{Nuovo Cimento} \textbf{52A},
122 (1967), Appendix E.
\bibitem{Kang:1975gt}K.~Kang and B.~Nicolescu, \emph{Phys. Rev.} \textbf{D 11},
2461 (1975).
\bibitem{Amaldi:1980kd}U.~Amaldi and K.~R.~Schubert, \emph{Nucl. Phys.}
\textbf{B166},
301 (1980).
\bibitem{log parametrisations}G. B. Yodh, Y. Pal and J. S. Trefil, \emph{Phys.
Rev. Lett.} \textbf{28},
1005 (1972); E. Leader and U. Maor, \emph{Phys. Lett.} \textbf{43B},
505 (1973); C. Bourrely and J. Fischer, \emph{Nucl. Phys.} \textbf{B61},
513 (1973); L. Lukaszuk and B. Nicolescu, \emph{Nuovo Cim. Lett.}
\textbf{8}, 405 (1973); K. Kang and A.R. White, \emph{Phys. Rev.} \textbf{D42},
835 (1990); M.~M.~Block, K.~Kang and A.~R.~White, \emph{Int. J.
Mod. Phys.} \textbf{A} \textbf{7}, 4449 (1992).
\bibitem{Pom}I. Ya. Pomeranchuk, \emph{Sov. Phys. JETP} \textbf{7}, 499 (1958) .
\bibitem{Cudell:2000tx}J.~R.~Cudell, V.~Ezhela, K.~Kang, S.~Lugovsky and
N.~Tkachenko,
\emph{Phys. Rev.} \textbf{D 61}, 034019 (2000) {[}hep-ph/9908218{]}.
Erratum-\emph{ibid.} \textbf{D63}, 059901 (2001); COMPAS Group contributions
in {\it Reviews of Particle Physics}, D. Groom et al., \emph{The European
Physical Journal} \textbf{C15}, 1 (2000) .
\bibitem{Dersch:2000zg}U.~Dersch \textit{et al.} {[}SELEX Collaboration{]},
\emph{Nucl. Phys.}
\textbf{B579}, 277 (2000) {[}hep-ex/9910052{]}.
\bibitem{Abbiendi:2000sz}G.~Abbiendi \textit{et al.} {[}OPAL Collaboration{]},
\emph{Eur. Phys. J.}
\textbf{C14}, 199 (2000) {[}hep-ex/9906039{]}.
\bibitem{Acciarri}M. Acciarri \emph{et al.} {[}L3 Collaboration{]},
hep-ex/0102025.
%%CITATION = HEP-EX 0102025;%%
\bibitem{Groom:2000in}D.~E.~Groom \textit{et al.}, \emph{Eur. Phys. J.}
\textbf{C15},
1 (2000).
\bibitem{Landshoff:2000mu}P.~V.~Landshoff, hep-ph/0010315; A.~Donnachie and
P.~V.~Landshoff,
\emph{Phys. Lett.} \textbf{B296}, 227 (1992) {[}hep-ph/9209205{]}%%CITATION =
%%HEP-PH 9209205;%%.
\bibitem{CKK}J. R. Cudell, K. Kang and S. K. Kim, \emph{Phys. Lett.}
\textbf{B395},
311 (1997) {[}hep-ph/9601336{]}. %%CITATION = HEP-PH 9601336;%%
\bibitem{Gauron:2000ri}P.~Gauron and B.~Nicolescu, \emph{Phys. Lett.}
\textbf{B486}, 
71 (2000){[}hep-ph/0004066{]}; B.~Nicolescu, \emph{Nucl. Phys. Proc. Suppl.}
\textbf{99}, 47 (2001) {[}hep-ph/0010292{]}.
\bibitem{solovev:1973}L.~D.~Soloviev, \emph{Pisma v ZHETF} \textbf{18},
455 (1973); \emph{Pisma
v ZHETF} \textbf{19}, 185 (1974) (in Russian); \emph{JETP Lett.} \textbf{19},
116 (1974), L.~D.~Soloviev and A.~V.~Shchelkachev, \emph{Fiz.
Elem. Chast. Atom. Yadra} \textbf{6}, 571-600 (1975) (in Russian),
\emph{Sov. J. Part. Nucl.} \textbf{6}, 229 (1975); M.~S.~Dubovikov
and K.~A.~Ter-Martirosian, \emph{Nucl. Phys.} \textbf{B124},
163 (1977); M.~S.~Dubovikov, B.~Z.~Kopeliovich, L.~I.~Lapidus and
K.~A.~Ter-Martirosian,
\emph{Nucl. Phys.} \textbf{B123}, 147 (1977); M.~S.~Dubovikov and
K.~A.~Ter-Martirosian, \textit{\emph{in}} \textit{Coral Gables 1976,
Proceedings, New Pathways In High-energy Physics, Vol.Ii, New York
1976, 313-329}.
\bibitem{Ter-Martirosian:1988yy}S.~S.~Gershtein and A.~A.~Logunov, Sov. J.
Nucl. Phys. \textbf{39},
960 (1984); \emph{Yad. Fiz.} \textbf{39}, 1514-1516 (1984), IFVE-84-37,
Feb 1984 (in Russian); Yu.~D.~Prokoshkin, \emph{Yad. Fiz.} \textbf{40},
1579-1584 (1984) (in Russian); K.~A.~Ter-Martirosian, \emph{Nucl. Phys.}
\textbf{A477}, 696 (1988).
\bibitem{Desgrolard}P. Desgrolard and E. Martynov, hep-ph/0105277. %%CITATION =
%%HEP-PH 0105277;%%
\bibitem{Levin:1965mi}E.~M.~Levin and L.~L.~Frankfurt, \emph{Pisma v ZHETF}
\textbf{3},
 652 (1965) (in Russian), E.~M.~Levin and L.~L.~Frankfurt, \emph{JETP
Lett.} \textbf{2}, 65 (1965).
\bibitem{Kiselev:1986fr}
A.~V.~Kiselev and V.~A.~Petrov,
%``Gluonic Structure And Strong Interactions Of Hadrons At High-Energies,''
\emph{Yad. Fiz.}  {\bf 44}, 1047 (1986)
[\emph{Sov. J. Nucl. Phys.}  {\bf 44}, 677 (1986)];
%%CITATION = YAFIA,44,1047;%%

\bibitem{jt}K.~Johnson and S.B.~Treiman, 
{\emph Phys. Rev. Lett.} {\bf 14 }, 189 (1965)
\bibitem{freund}P.~G.~O.~Freund, {\emph Phys. Rev. Lett. }{\bf 15 }, 929 (1965)
\bibitem{Feigenbaum:1997gy}J.~A.~Feigenbaum, P.~G.~Freund and M.~Pigli,
\emph{Phys. Rev.}
\textbf{D 56}, 2596 (1997) {[}hep-ph/9703296{]}.
\bibitem{Lipkin:1999tc}H.~J.~Lipkin, \emph{Phys. Rev.} \textbf{D11},
1827 (1975); H.~J.~Lipkin,
hep-ph/9911259.
\bibitem{Baltrusaitis:1984ka}R.~M.~Baltrusaitis \textit{et al.}, \emph{Phys.
Rev. Lett.} \textbf{52},
 1380 (1984).
\bibitem{Honda:1993kv}M.~Honda \textit{et al.}, \emph{Phys. Rev. Lett.}
\textbf{70},
525 (1993).
\bibitem{Durand:1988cr}L.~Durand and H.~Pi, \emph{Phys. Rev.} \textbf{D 38},
78 (1988).
\bibitem{Kopeliovich:1989iy}B.~Z.~Kopeliovich, N.~N.~Nikolaev and
I.~K.~Potashnikova, \emph{Phys. Rev.}
\textbf{D 39}, 769 (1989).
\bibitem{Nikolaev:1993mc}N.~N.~Nikolaev, \emph{Phys. Rev.} \textbf{D 48},
1904 (1993) {[}hep-ph/9304283{]}.
\bibitem{Velasco:1999ce}J.~Velasco, J.~Perez-Peraza, A.~Gallegos-Cruz,
M.~Alvarez-Madrigal,
A.~Faus-Golfe and A.~Sanchez-Hertz, in \emph{Salt Lake City 1999},
Cosmic ray, vol. 1, 198-201 {[}hep-ph/9910484{]}.
\bibitem{Block:2000pg}M.~M.~Block, F.~Halzen and T.~Stanev, \emph{Phys. Rev.}
\textbf{D
62}, 077501 (2000) {[}hep-ph/0004232{]}.
\bibitem{Desgrolard:2001sf}P.~Desgrolard, M.~Giffon, A.~Lengyel and
E.~Martynov, \emph{Nuovo
Cim.} \textbf{A107}, 637 (1994); P.~Desgrolard, M.~Giffon, E.~Martynov
and E.~Predazzi, \emph{Eur. Phys. J.} C \textbf{18}, 555 (2001)
{[}hep-ph/0006244{]}.
\bibitem{log2 very high E}L.L. Jenkovszky, B.V. Struminsky, A.N. Wall,
\emph{Yad. Fiz.} {\bf 46},
1519 (1986); A.N. Wall, L.L. Jenkovszky and B.V. Struminsky, \emph{Fiz.
Elem. Chast. Atom. Yadra} {\bf 19}, 180 (1988) (In Russian). %%CITATION =
%%FECAA,19,180;%%
\bibitem{FFKT}J. Finkelstein, H.M. Fried, K. Kang and C-I. Tan, \emph{Phys.
Lett.}
{\bf B 232}, 257 (1989).
\bibitem{Gauron 88}P. Gauron, B. Nicolescu and E. Leader, \emph{Nucl. Phys.}
\textbf{B
299}, 640 (1988). %%CITATION = NUPHA,B299,640;%%
\bibitem{Bartels:2000yt}J.~Bartels, L.~N.~Lipatov and G.~P.~Vacca, \emph{Phys.
Lett.}
\textbf{B477}, 178 (2000)  {[}hep-ph/9912423{]}; G.~P.~Vacca, \emph{Phys.
Lett}\textbf{\emph{.}}
\textbf{B489}, 337 (2000) {[}hep-ph/0007067{]}.
\bibitem{Cudell 2001}J.R. Cudell and G. Soyez, hep-ph/0106307. %%CITATION =
%%HEP-PH 0106307;%%
\bibitem{Haitun}S. D. Haitun, \emph{Scientometrics} {\bf 10}, 3 and 133 (1986),
\emph{Scientometrics}
{\bf 15}, 45 (1989).
\bibitem{COMPETE}See the preliminary version of a Web interface at the address
http://sirius.ihep.su/\textasciitilde{}kuyanov/OK/eng/intro.html.
It is planned to make the detailed numerical data resulting from the
fits easily available on the Web in a computer-readable form. Meanwhile
these data can be obtained by request from tkachenkon@mx.ihep.su (with
CC to: kuyanov@mx.ihep.su,ezhela@mx.ihep.su).\vspace{11cm}
\end{thebibliography}
\end{document}